\documentclass[fleqn,usenatbib]{mnras}

\usepackage{newtxtext,newtxmath}
\usepackage{booktabs}

\usepackage[T1]{fontenc}

\DeclareRobustCommand{\VAN}[3]{#2}
\let\VANthebibliography\thebibliography
\def\thebibliography{\DeclareRobustCommand{\VAN}[3]{##3}\VANthebibliography}

\usepackage{graphicx}	
\usepackage{amsmath}	

\usepackage{orcidlink}

\title[Atmospheric Characterisation of WASP-96\,b]{Awesome SOSS: Atmospheric Characterisation of WASP-96\,b using the JWST Early Release Observations}

\author[Taylor et al.]{Jake Taylor\orcidlink{0000-0003-4844-9838}$^{1,2}$\thanks{E-mail: jake.taylor@physics.ox.ac.uk},
Michael Radica\orcidlink{0000-0002-3328-1203}$^{2}$,
Luis Welbanks\orcidlink{0000-0003-0156-4564}$^{3}$\thanks{NASA Sagan Fellow},
Ryan J.\ MacDonald\orcidlink{0000-0003-4816-3469}$^{4}$\thanks{NASA Sagan Fellow},
Jasmina Blecic\orcidlink{0000-0002-0769-9614}$^{5, 6}$
\newauthor
Maria Zamyatina\orcidlink{0000-0002-9705-0535}$^{7}$,
Alexander Roth$^{1}$,
Jacob L.\ Bean\orcidlink{0000-0003-4733-6532}$^{8}$,
Vivien Parmentier$^{1,9}$,
Louis-Philippe Coulombe\orcidlink{0000-0002-2195-735X}$^{2}$,
\newauthor
Adina D.\ Feinstein\orcidlink{0000-0002-9464-8101}$^{8}$,
Néstor Espinoza\orcidlink{0000-0001-9513-1449}$^{10,11}$,
Björn Benneke$^{2}$,
David Lafrenière\orcidlink{0000-0002-6780-4252}$^{2}$, 
René Doyon\orcidlink{0000-0001-5485-4675}$^{2}$
\newauthor
and Eva-Maria Ahrer\orcidlink{0000-0003-0973-8426}$^{12,13}$
\\
$^{1}$ Department of Physics (Atmospheric, Oceanic and Planetary Physics), University of Oxford, Parks Rd, Oxford OX1 3PU, UK\\
$^{2}$ Institut Trottier de Recherche sur les Exoplanètes and Département de Physique, Université de Montréal, 1375 Avenue Thérèse-Lavoie-Roux,\\ Montréal, QC, H2V 0B3, Canada \\
$^{3}$ School of Earth and Space Exploration, Arizona State University, 781 Terrace Mall, Tempe, AZ, 85287, USA\\
$^{4}$ Department of Astronomy, University of Michigan, 1085 S. University Ave., Ann Arbor, MI 48109, USA\\
$^{5}$ Department of Physics, New York University Abu Dhabi, PO Box 129188 Abu Dhabi, UAE\\
$^{6}$ Center for Astro, Particle, and Planetary Physics (CAP3), New York University Abu Dhabi, PO Box 129188 Abu Dhabi, UAE\\
$^{7}$ Department of Physics and Astronomy, Faculty of Environment, Science and Economy, University of Exeter, Exeter EX4 4QL, UK \\
$^{8}$ Department of Astronomy \& Astrophysics, University of Chicago, 5640 S Ellis Ave, Chicago, IL 60637, USA\\
$^{9}$ Université Côte d’Azur, Observatoire de la Côte d’Azur, CNRS, Laboratoire Lagrange, France \\
$^{10}$ Space Telescope Science Institute, 3700 San Martin Drive, Baltimore, MD 21218, USA\\
$^{11}$ Department of Physics and Astronomy, Johns Hopkins University, 3400 N Charles St, Baltimore, MD 21218, USA\\
$^{12}$Centre for Exoplanets and Habitability, University of Warwick, Coventry, UK \\
$^{13}$ Department of Physics, University of Warwick, Coventry, UK \\
}

\date{Accepted XXX. Received YYY; in original form ZZZ}

\pubyear{2022}

\begin{document}
\label{firstpage}
\pagerange{\pageref{firstpage}--\pageref{lastpage}}
\maketitle

\begin{abstract}
The newly operational JWST offers the potential to study the atmospheres of distant worlds with precision that has not been achieved before. One of the first exoplanets observed by JWST in the summer of 2022 was WASP-96\,b, a hot-Saturn orbiting a G8 star. As part of the Early Release Observations program, one transit of WASP-96\,b was observed with NIRISS/SOSS to capture its transmission spectrum from 0.6--2.85\,µm. In this work, we utilise four retrieval frameworks to report precise and robust measurements of WASP-96\,b's atmospheric composition. We constrain the logarithmic volume mixing ratios of multiple chemical species in its atmosphere, including: H$_2$O = $-3.59 ^{+ 0.35 }_{- 0.35 }$, CO$_2$ = $-4.38 ^{+ 0.47 }_{- 0.57 }$ and K = $-8.04 ^{+ 1.22 }_{- 1.71 }$. Notably, our results offer a first abundance constraint on potassium in WASP-96\,b's atmosphere, and important inferences on carbon-bearing species such as CO$_2$ and CO. Our short wavelength NIRISS/SOSS data are best explained by the presence of an enhanced Rayleigh scattering slope, despite previous inferences of a clear atmosphere --- although we find no evidence for a grey cloud deck. Finally, we explore the data resolution required to appropriately interpret observations using NIRISS/SOSS. We find that our inferences are robust against different binning schemes. That is, from low $R = 125$ to the native resolution of the instrument, the bulk atmospheric properties of the planet are consistent. Our systematic analysis of these exquisite observations demonstrates the power of NIRISS/SOSS to detect and constrain multiple molecular and atomic species in the atmospheres of hot giant planets.  
\end{abstract}

\begin{keywords}
planets and satellites: atmospheres -- planets and satellites: gaseous planets -- planets and satellites: individual: WASP-96\,b
\end{keywords}



\section{Introduction}
After launch in December 2021, a careful journey to L2, and a successful commissioning period, JWST finally began its long-awaited science operations on July 12, 2022. It is a credit to how far the field of exoplanet astronomy has progressed in the past couple of decades that some of the very first observations with this revolutionary new observatory were of transiting exoplanets. JWST vastly extends the wavelength range with which exoplanet atmospheres can be probed from space. Previous state-of-the-art observations with the Hubble Space Telescope (HST) probed UV, optical, and near infrared wavelengths out to 1.7\,µm. The Spitzer Space Telescope enabled predominantly photometric measurements further into the infrared, although only the bluest bandpasses at 3.6 and 4.5\,µm remained in operation after the coolant ran out in 2009. In combination, the four instruments on board the JWST allows for the study of exoplanet atmospheres from 0.6--28\,µm; enabling the characterization of their atmospheres at wavelengths never seen before, broadening our discovery space into uncharted territories. This is evident from the first results from the Early Release Science (ERS) observations of the hot-Jupiter WASP-39\,b which yielded the first ever detections of CO$_2$ and SO$_2$ \citep{ERS2022,Alderson2022,Rustamkulov2022,Tsai2022}.


The Early Release Observations (ERO) program was designed to provide the astronomical community with publicly available data, touching on many of the key science objectives of JWST, immediately after the end of the commissioning period \citep{Pontoppidan2022}. For the exoplanet portion of the ERO program, transits of two hot-Saturns, WASP-96\,b \citep{Hellier2014} and HAT-P-18\,b \citep{Hartman2011, Fu2022} were observed with the Single Object Slitless Spectroscopy (SOSS) mode (Albert et al.~submitted) of the Near Infrared Imager and Slitless Spectrograph (NIRISS) instrument (Doyon et al.~submitted).

This work focuses on WASP-96\,b, an inflated hot-Saturn exoplanet with a mass of 0.498$\pm$ 0.03\,M$\rm _J$, a radius of 1.2$\pm$0.06\,R$\rm _J$, and an equilibrium temperature of $\sim$1300\,K. It orbits a G8 star in the constellation of Phoenix with an orbital period of 3.4\,d. The planet's short orbital period, combined with its low density makes it an ideal candidate for atmospheric spectroscopy. Indeed, there have been multiple previous atmospheric studies of this planet, with the first being a ground-based spectrum using VLT/FORS2 by \citet{Nikolov2018}. This observation covered a spectral range of 0.36--0.82\,µm using spectroscopic bins with widths of 0.016\,µm. This spectroscopic precision allowed for the measurement of the pressure-broadened sodium D line with wings reported to cover six atmospheric pressure scale heights \citep{Nikolov2018}. The visibility of the sodium wings suggested that there are no clouds or hazes obscuring them in the atmosphere at the pressure ranges probed in transmission \citep{Fortney2005}. This was supported by their atmospheric modeling which finds no evidence for additional opacity due to clouds. \citet{Nikolov2018} further conclude that the abundance of Na is consistent with the measured stellar value. 

Prior to the commissioning period of JWST, \citet{Nikolov2022} published the transmission spectrum of WASP-96\,b using HST and Spitzer, providing the first look at the infrared spectrum of the planet using space-based instrumentation. To explore the atmospheric constituents in detail they couple the HST and Spitzer observations with the previous VLT observations. They find an offset between the space and ground-based data, consistent with what was found by \citet{Yip2021} who explored the impact of combining space-based and ground-based observations. Together, the HST and Spitzer observations confirm previous findings that the transmission spectrum is consistent with a cloud-free atmosphere. They are able to put a constraint on the absolute sodium and oxygen abundance and find them to be 21$^{+27}_{-14}$ $\times$ and 7$^{+11}_{-4}$ $\times$ solar values, respectively.

Later, \citet{McGruder2022} published a study of WASP-96\,b adding to the existing ensemble of transit measurements using the ground-based telescope IMACS/Magellan as part of the ACCESS project \citep{Jordan2013,Rackham2017}. Their transmission spectrum covers the spectral range of 0.44--0.9\,µm, which overlaps with the VLT/FORS2 observations of \citet{Nikolov2018}, enabling an independent confirmation of the sodium feature with its pressure-broadened wings. They combine their two transits with the published VLT/FORS2 and HST data and perform spectral retrievals on the combined spectrum (0.4--1.644\,µm). Their results indicate solar-to-super solar abundances of Na and H$_2$O, with log-mixing ratios of $-5.4^{+2.0}_{-1.9}$ and $-4.5^{+2.0}_{-2.0}$ respectively, in rough agreement with \citet{Nikolov2022} and with previous suggestions of super-solar alkali abundances \citep{Welbanks2019}.

This study is the second paper in a two-part series providing an in-depth treatment of the ERO observations of WASP-96\,b. The companion paper, Radica et al.~(submitted) focuses on the reduction and extraction of the planet's transmission spectrum from the SOSS time series observations (TSO), as well as provides some initial insights into the composition of WASP-96\,b's atmosphere through comparisons of the planet's transmission spectrum with grids of self-consistent atmospheric models. They conclude that the NIRISS/SOSS observations of WASP-96\,b are best explained by a cloud-free atmosphere, with a solar-to-super-solar metallicity atmosphere and solar carbon-to-oxygen ratio (C/O). In this study, we perform a detailed atmospheric characterization of WASP-96\,b using the spectrum presented in Radica et al.~(submitted). NIRISS/SOSS provides spectral coverage from 0.6--2.8\,µm, covering the red wing of the 0.59\,µm sodium doublet as well as multiple water bands. Hence we can assess the robustness of previous observations \citep{Nikolov2022,McGruder2022} and independently confirm them with an instrument specifically built to study exoplanet atmospheres. 

In the era of HST, exoplanet atmosphere observations were generally binned to a resolution which achieved a sufficient signal to noise ratio to obtain spectral information. The choice of binning, though, has remained somewhat arbitrary; for example, with HST/WFC3 G141 (1.1--1.7 microns), \citet{Lineb2016} use 10 spectral bins for the secondary eclipse of HD 209458b and \citet{Kreidberg2014} use 22 spectral bins for the transmission spectrum and 15 spectral bins for the secondary eclipse of WASP-43b. Hence, for the same instrument, different spectral binning was used throughout the literature, and as of yet, there has been no exploration of whether different spectral bin choices impact the inferred atmospheric properties. 

We therefore aim to answer the following questions:
\begin{enumerate}
    \item What are the chemical species present in WASP-96\,b's atmosphere are what are their abundances? Is the atmosphere indeed cloud free?
    \item How robust against the choice of framework and model assumptions are the retrieved chemical abundances?
    \item Does binning the data from native to lower resolution produce different inferred abundances?
\end{enumerate}

This work is organized as follows: we provide a brief overview of the observations and data reduction in Section~\ref{sec: Observations}. We outline the different modelling strategies, both through inference (e.g., retrievals) and forward models, in Section~\ref{sec: Spectral Analysis}, and present the modelling results in Section~\ref{sec: Results}. Section~\ref{sec: Discussion} contains a brief discussion of these results, and we summarize our work in Section~\ref{sec: Conclusions}.

\section{Observations}
\label{sec: Observations}

One transit of the hot-Saturn WASP-96\,b was observed with NIRISS/SOSS on June 21, 2022 as part of the JWST Early Release Observations (ERO) program \citep{Pontoppidan2022}. The total duration of the time series observation (TSO) was 6.4\,hr. The SUBSTRIP256 subarray configuration was used to capture the first three diffraction orders of the target star on the detector (Albert et al.~submitted), which provides access to the full 0.6--2.8\,µm wavelength range of the SOSS mode. Order 1 covers the 0.85--2.8\,µm wavelength range, with the 0.6--0.85\,µm information provided by order 2. The third order is generally too faint to be extracted, and does not provide any unique wavelength coverage (Albert et al.~submitted).

\begin{figure*}
    \centering
    \includegraphics[width=0.99\textwidth]{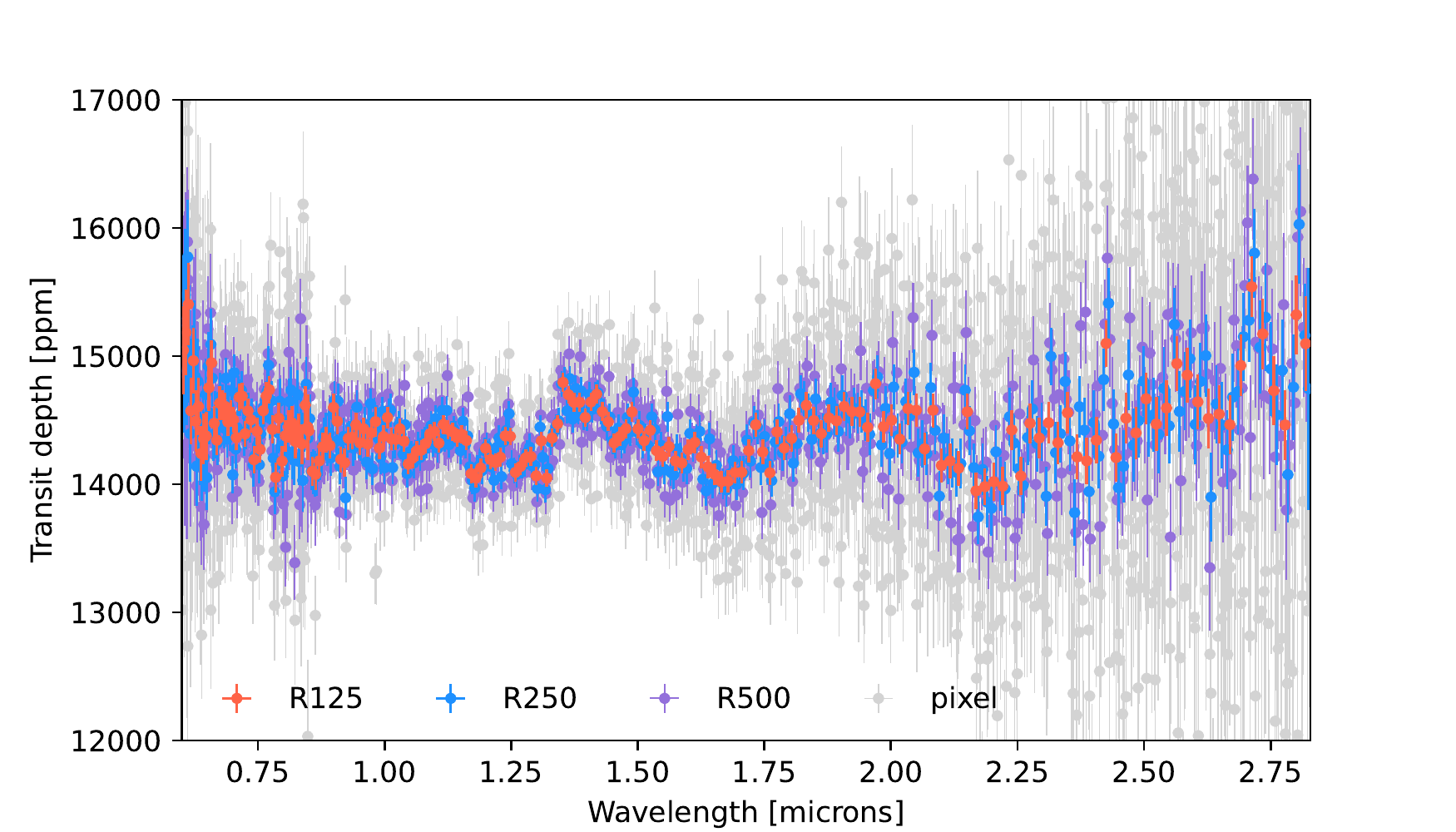}
    \caption{The full 0.6--2.8\,µm NIRISS/SOSS spectrum of WASP-96\,b from Radica et al.~(submitted). The pixel level spectrum is shown in grey, as well as when binned to a resolution of R=500, 250, and 125 in purple, blue, and red respectively.}
    \label{fig:binned_spectra}
\end{figure*}

The reduction of these data, using the \texttt{supreme-SPOON} pipeline\footnote{\url{https://github.com/radicamc/supreme-spoon}} \citep[][Radica et al.~submitted]{feinsteinw39, CoulombeW18b} are treated in depth in Radica et al.~(submitted). In that work, they present a walkthrough of the critical reduction steps, including correction of the zodiacal background light and 1/$f$ noise. Their final transmission spectrum, which we make use of in this work, was extracted with the \texttt{ATOCA} algorithm \citep{darveau-bernier_atoca, radica_applesoss} to explicitly model the self-contamination of the first and second diffraction orders on the detector. The transit depths were additionally fit at the pixel level (that is, one transit depth per pixel column on the detector), and were post-processed to correct for contamination from background field stars, which can occur due to the slitless nature of the SOSS mode. This spectrum is shown at the pixel-level, as well as binned several lower resolutions in Figure~\ref{fig:binned_spectra}. Their spectrophotometric light curves reach an average precision of 1.2$\times$ and 1.4$\times$ the photon noise for order 1 and 2 respectively, resulting in an average pixel-level transit depth precision of 522\,ppm and 534\,ppm respectively. 

\section{Spectral Analysis}
\label{sec: Spectral Analysis}

We use two different modelling approaches to thoroughly explore WASP-96\,b's atmosphere. The first is comparisons with forward models computed using 3D General Circulation Models (GCMs). These allow us to explore the potential formation of different species of cloud condensates in the atmosphere of WASP-96\,b \citep[e.g.,][]{Samra2022}, and also consider chemical kinetics \citep{Zamyatina2023}. The second is a spectral retrieval analysis that allows us to infer the atmospheric properties of WASP-96\,b, such as its chemical composition and temperature structure, directly from the Radica et al.~(submitted) transmission spectrum (Figure~\ref{fig:binned_spectra}). For the retrieval analysis, we use four different codes: CHIMERA \citep{Line2013}, Aurora \citep{Welbanks2021}, POSEIDON \citep{MacDonald2017,MacDonald2023} and PyratBay  \citep{CubillosBlecic2021MNRAS} to ensure that the atmospheric composition we retrieve is robust against the choice of retrieval code. The model setups for each of the two approaches are detailed below.

\subsection{3D Forward Modelling using GCMs}
\label{sec: 3D Forward Modelling}

Although all previous observational studies of this planet have concluded a cloud-free upper atmosphere for WASP-96\,b \citep{Nikolov2018, Nikolov2022, McGruder2022}, the idea that clouds could be present in the atmosphere of this planet was recently explored by \cite{Samra2022}. Their 3D GCM model considers a kinetic, non-equilibrium formation model for mixed material cloud particles. Their GCM models show that clouds could indeed be ubiquitous in the low-pressure, terminator regions of WASP-96\,b's atmosphere, with silicate and metal oxide clouds being the most prominent condensate species. They conclude that the \citet{Nikolov2022} transmission spectrum can also be fit with cloudy models. However, whether the clouds predicted by the kinetic model actually form in WASP-96\,b depends on whether they are cold-trapped below the photosphere~\citep{Parmentier2016,Powell2018}, a mechanism that cannot currently be resolved with the kinetics models.


We perform our own GCM modelling to investigate the plausibility that WASP-96\,b could host clouds. For our first GCM analysis, we use the non-grey SPARC/MITgcm \citep{2009ApJ...699..564S}. Specifically, we make use of the large grid of models generated by Roth et al.~(in prep). The GCM setup is very similar to that described in \citet{2018A&A...617A.110P,2021MNRAS.501...78P}, but does not consider cloud condensation. The model grid spans a wide range of equilibrium temperatures, atmospheric metallicity, orbital period, and surface gravities which is then interpolated into the specific WASP-96\,b parameters. However, other parameters such as the planet radii are fixed and the models have an infinite drag timescale. The resulting thermal profiles are then interpolated to the system parameters of WASP-96\,b.  

\begin{figure*}
    \centering
    \includegraphics[width=1.\textwidth]{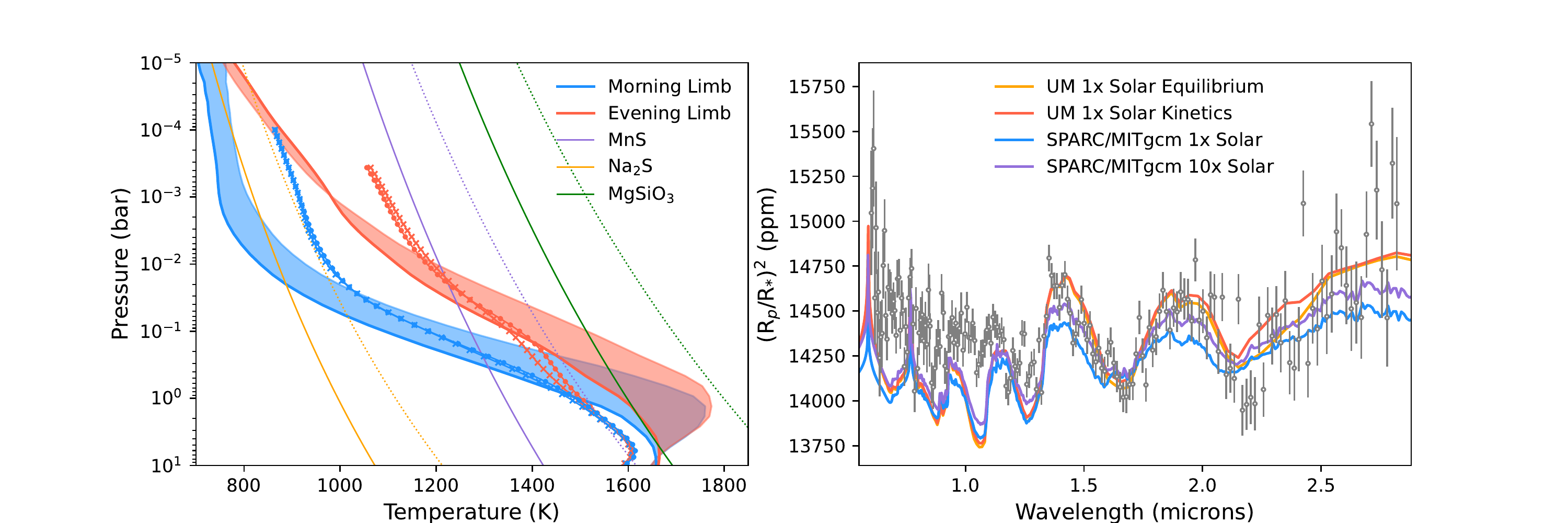}
    \caption{
    \emph{Left:} Temperature profiles generated from the SPARC MIT/gcm and the UM. The morning and evening limbs are shown in blue and red respectively. The solid blue line is the 1$\times$ solar model from the MIT/gcm, with the shading showing the parameter space covered between a 1$\times$ solar and 10$\times$ solar model, both assuming chemical equilibrium. The crosses and circles are from the UM, showing equilibrium and kinetics cases respectively, both for a 1x solar metallicity. Condensation curves for three different cloud species are shown: Na$_2$S, MnS, and MgSiO$_3$ in orange, purple, and green respectively. The line style denotes curves for different atmospheric metallicities: solid, dashed and dotted for 1$\times$ and 10$\times$ solar, respectively. 
    \emph{Right:} WASP-96\,b NIRISS/SOSS transmission spectrum from Radica et al.~(submitted) binned to a resolution of R=125 (black points with error bars) compared to simulated transmission spectra from outputs of the UM and SPARC MIT/gcm. The orange and red lines are UM models, with and without considering kinetics. The blue and purple lines are from the SPARC MIT/gcm 1$\times$ and 10$\times$ solar metallicity runs, respectively.}
    \label{fig:transmission_spectrum_gcm}
\end{figure*}

The thermal profiles are read into CHIMERA (see Section~\ref{subsubsec:CHIMERA} for more details) to produce a transmission spectrum of an atmosphere that has a solar C/O and metallicity of 1$\times$ and 10$\times$ solar. The thermal profiles and transmission spectra are shown in Figure~\ref{fig:transmission_spectrum_gcm}. To capture the PT profile parameter space spanned by our range of considered metallicities, we denote the 1$\times$ solar profile in a solid line, and the edge of the shaded area denotes the 10$\times$ solar profile. It can be seen that the PT structures cross the condensation curves for various cloud species. Specifically, the SPARC/MITgcm predicts that the morning limb favours high-altitude Na$_2$S clouds with deeper MnS clouds, whereas in the evening limb only MnS clouds could condense in the pressure regions probed by transmission. Silicate clouds should form only in the 1$\times$ solar metallicity case, and only in the deep layers of the atmosphere ($\sim$1\,bar). Depending on whether vertical mixing is large enough, they could be efficiently mixed up to the pressure levels probed by the observations or remain trapped in the deep layers of the atmosphere~\citep{Powell2018}. 

Our second GCM analysis utilizes the Met Office \textsc{Unified Model} (\textsc{UM}) run specifically for WASP-96\,b. We used the same basic model setup as in \citet{Drummond2020} and \citet{Zamyatina2023} with the following changes: (a) the \textsc{PHOENIX} BT-Settl stellar spectrum \citep{Rajpurohit2013} closely matching WASP-96, (b) WASP-96\,b parameters from \citet{Hellier2014}, and (c) the \textsc{UM} version 11.6, initialised with (d) WASP-96\,b dayside-average pressure-temperature profile obtained with the 1D radiative-convective-chemistry model \textsc{ATMO} \citep{Drummond2016} assuming chemical equilibrium for the chemical species present in the \citet{Venot2012} chemical network. We further assume the atmosphere to be cloud/haze free and have a solar metallicity and C/O ratio based on the initial modelling of Radica et al.~(submitted). 

Within the \textsc{UM} framework, we ran two simulations, each with a different chemical scheme: one assuming chemical equilibrium, and the other a chemical kinetics scheme, which computes the production and loss of the chemical species present in the \citet{Venot2019} reduced chemical network. We will refer to these simulations as "UM 1$\times$ solar equilibrium" and "UM 1$\times$ solar kinetics", respectively, with the latter simulation accounting for the opacity changes not only due to changes in pressure and temperature, but also due to the transport of chemical species in the atmosphere. The left panel of Figure~\ref{fig:transmission_spectrum_gcm} shows that both \textsc{UM} simulations predict similar limb-average PT profiles (weighted over all latitudes and $\pm20^{\circ}$ longitude), with the morning limb being colder than the evening limb at pressures $<$1 bar. Contrary to the SPARC/MITgcm, pressure-temperature profiles from the UM suggest that only MnS clouds could form on WASP-96\,b's limbs. This is because the \textsc{UM} predicts a shallower temperature gradient at pressures $<$10$^{-2}$ bar causing the \textsc{UM} to have temperatures 100--200\,K higher than those predicted at comparable pressures by the SPARC/MITgcm. Both GCMs predict similar positions for the MnS cloud decks on both limbs, when the assumed metallicity is 1$\times$ solar. However, given that the MnS nucleation rate is relatively low \citep{Gao2020}, these clouds might not form quickly enough for their opacity to be relevant for WASP-96\,b.

The right panel of Figure~\ref{fig:transmission_spectrum_gcm} shows that both the \textsc{UM} and the SPARC/MITgcm simulations produce transmission spectra that agree well with WASP-96\,b's JWST NIRISS/SOSS transmission spectrum in the range 1.3--2.15\,µm. Blueward of 1.3\,µm, however, K and H$_2$O features are muted relative to those predicted by the haze and cloud free GCM models, suggesting the presence of a scattering opacity source. Redward of 2.15\,µm, the observations and the models broadly agree, but the observed transit depths vary highly with wavelength. Of particular note is the region between 2.15--2.5\,µm, where the "UM 1$\times$ solar kinetics" simulation predicts a higher transit depth than the "UM 1$\times$ solar equilibrium" simulation. This difference is caused by an enhancement of the abundance of CH$_4$ due to transport-induced quenching, which is captured only in the UM kinetics simulation. However, we are not able to robustly distinguish between these two cases with the current data. Another difference is that, because the solar-composition SPARC/MITgcm predicts cooler limb temperatures than the \textsc{UM} simulation, the spectral features of the SPARC/MITgcm are shallower than the observations. However, the SPARC/MITgcm 10$\times$ solar metallicity leads to a hotter thermal profile and thus to a better match to the data. This difference highlights the intrinsic dependence of the observables to the modelling framework when using complex, 3D, GCMs \citep{Showman2020}. 

Overall, both clear-sky GCMs used in this study provide good agreement with our JWST NIRISS/SOSS transmission spectrum of WASP-96\,b. However, we are not able to robustly distinguish between 1$\times$ and 10$\times$ solar metallicity models with the current data, and both models struggle to reproduce the observations blueward of 1.3\,µm. This further motivates an in-depth investigation using atmospheric retrievals.

\subsection{Atmospheric Retrieval}
\label{subsec:retrievals}

Atmospheric retrievals are a powerful tool to extract information about an exoplanet atmosphere directly from the data \citep{Madhusudhan2009}. We explore the data in a hierarchical way, from simple (e.g., cloud free, free abundances, isothermal) to complex models (e.g., inclusion of hazes and clouds, chemical equilibrium, non-isothermal), with multiple retrieval codes. The first set of retrievals we perform are `free chemistry' retrievals, which directly infer the volume mixing ratios (VMR) for a set of chemical species assumed to be present in the atmosphere (the VMRs are assumed constant with altitude). Each retrieval framework assumed that the atmosphere is dominated by H$_2$ -- expected for objects that have physical properties similar to Saturn, and included the same molecules as opacity sources. All frameworks use the WASP-96 system parameters reported in \citet{Hellier2014}. The second set of retrievals are performed assuming that the vertical abundances of the chemical species are in thermochemical equilibrium.

To robustly interpret our WASP-96\,b observations, we employ four different retrieval frameworks: CHIMERA \citep{Line2013}, Aurora \citep{Welbanks2021}, POSEIDON \citep{MacDonald2017,MacDonald2023} and PyratBay \citep{CubillosBlecic2021MNRAS}. A multiple-retrieval approach allows us to compare our results in the regime of high-precision data \citep{Barstow2020,Barstow2022}, thereby quantifying the stability of our atmospheric inferences to model implementations. The common molecules to each code are: H$_2$O, CO, CO$_2$, CH$_4$, NH$_3$, HCN, Na and K, these have a prior U($-12$,$-1$)\footnote{with the exception of Aurora which has U($-12$,$-0.3$) \citep{Welbanks2019}} for all VMRs. The set-up of each code is explained in the following subsections. Furthermore, CHIMERA is also used to run a chemical equilibrium retrieval, as an additional test.

\subsubsection{CHIMERA}
\label{subsubsec:CHIMERA}

We use CHIMERA\footnote{The open source code can be found here: \url{https://github.com/mrline/CHIMERA}} to perform both free and chemically consistent spectral retrievals. CHIMERA is the only framework in this study that uses the correlated-$k$ approach \citep{Lacis1991} when computing transmission through the atmosphere. The $k$-tables are computed at a resolution of R=3000; the line-by-line data used to calculate the $k$-tables are from the following sources: H$_2$O \citep{Polyansky2018, Freedman2014}, CO$_2$ \citep[][]{Freedman2014}, CO \citep{Rothman+10}, CH$_4$ \citep{Rothman+10}, HCN \citep{Barber2014},  Na \citep[]{Kramida2018, Allard2019}, and K \citep[]{Kramida2018, Allard+16}, and were computed following the methods described in \citet{Gharib2021, Grimm2021}. We assume the atmosphere is dominated by H$_2$, with a He/H$_2$ ratio of 0.1764; therefore, we also model the H$_2$-H$_2$ and H$_2$-He collision-induced absorption (CIA) \citep[][]{Richard2012}.

To compute the thermal structure, we use the parameterisation described in \cite{Madhusudhan2009}. This approach splits the atmosphere into three layers: the upper atmosphere, where no inversion can take place, a middle region, where an inversion is possible, and a deep layer, where the thermal structure is isothermal. We also consider a scenario in which the temperature structure is isothermal, and find that the abundances do not dependent on the thermal structure parameterisations. We also consider an atmosphere that is just parameterised by an isothermal model, it can be seen that all retrievals tend towards an isothermal temperature structure \ref{fig:WASP-96b_Results}.

Our chemically consistent retrievals aim to explore the impact of physical coupling between the atmospheric composition and temperature structure. Specifically, the molecular and atomic vertical abundances are assumed to be in thermochemical equilibrium. The equilibrium abundances are computed using the NASA CEA (Chemical Equilibrium with Applications) model \citep{gordon1994computer} for a given C/O, metallicity, and temperature structure. Thus, the C/O ratio and metallicity are free parameters for these retrievals instead of the chemical abundances themselves.

We model hazes following the prescription of \citet{Lecavelier2008}, which treats hazes as enhanced H$_2$ Rayleigh scattering with a free power law slope. This parameterisation expresses the opacity as $\sigma_{\text{Hazes}} = \alpha \sigma_0 (\lambda/\lambda_0)^{\gamma} $, where $\alpha$ is the Rayleigh enhancement factor and $\gamma$ is the scattering slope (equal to $-4$ for H$_2$ Rayleigh scattering). $\sigma_0$ is the H$_2$ Rayleigh cross section at $\lambda_0$, given by 2.3$\times$10$^{-27}$ cm$^2$ and 430 nm respectively. Alongside the haze calculation, we fit for a constant-in-wavelength grey cloud with opacity $\kappa_\text{cloud}$. Hence, we term this model ``Simple Haze + Cloud model".

To explore the parameter space, we coupled our parametric forward model with the Bayesian Nested Sampling algorithm \textsc{PyMultiNest} \citep{feroz09,buchner14}.

\subsubsection{Aurora}

We complement our atmospheric analysis by inferring the atmospheric properties of WASP-96\,b using Aurora \citet{Welbanks2021}, a Bayesian atmospheric retrieval framework for the interpretation of ground- and space-based observations of transiting exoplanets. Our atmospheric model setup generally follows a similar approach to previous atmospheric studies \citep[e.g.,][]{Welbanks2019a} with the same priors for WASP-96\,b as in the analysis of the existing VLT observations \citep{Nikolov2018} presented in \citet{Welbanks2019}. Our atmospheric model computes line-by-line radiative transfer in transmission geometry in a plane-parallel atmosphere. The pressure structure of the atmosphere assumes hydrostatic equilibrium for a varying-with-height gravity, in a grid of 100 layers uniformly distributed in log-pressure from $10^{-7}$ to 100 bar. The Bayesian inference is performed using the framework MultiNest \citep{feroz09} through its Python implementation PyMultiNest \citep{buchner14} using 2000 live points.

We explore a series of atmospheric model scenarios with Aurora including the possibility of multidimensional clouds and hazes \citep[e.g.,][]{Welbanks2019}, terminator inhomogeneities \citep[e.g.,][]{Welbanks2022}, and other modelling assumptions regarding the number of free parameters in our retrievals \citep[e.g.,][]{Welbanks2019a}. Through this exploration of models we determined a fiducial 19 parameter model for our `free retrieval' analysis using Aurora and other similar frameworks. This model setup considers a non-isothermal pressure-temperature structure parameterized using the six parameter prescription of \citet{Madhusudhan2009}. Eight sources of opacity are considered in our models. These species, expected to be the main absorbers for hot gas giants \citep[e.g.,][]{Madhusudhan2019}, are parameterized by their logarithmic volume mixing ratios assumed to be constant with height. The species and their corresponding line lists are CH$_4$ \citep{Yurchenko2014, Yurchenko2017}, CO \citep{Rothman2010}, CO$_2$ \citep{Rothman2010}, H$_2$O \citep{Rothman2010}, HCN \citep{Barber2014}, K \citep{Allard2016}, Na \citep{Allard2019}, and NH$_3$ \citep{YurchenkoEtal2011}. We further include H$_2$–-H$_2$ and H$_2$–He collision induced absorption \citep[CIA;][]{Richard2012} and H$_2$-Rayleigh scattering \citep{DalgarnoWilliams1962}. The opacities are computed following the methods described in \citet{Gandhi2017, Gandhi2018, Gandhi2020} and \citet{Welbanks2019}. 

We consider the presence of clouds and hazes in our atmospheric models using the modeling strategy for inhomogeneous terminator cover presented in \citet{Line2016}. We consider the presence of scattering hazes as deviations from the Rayleigh scattering in the models by following the parameterization of \citet{Lecavelier2008} as described above. The spectroscopic effect of clouds is included by considering the presence optically thick cloud decks at a specific pressure level. The combination of inhomogeneous clouds and hazes is implemented following the single-sector prescription as explained in \citet{Welbanks2019} using four additional free parameters. Finally, we use one free parameter to infer the reference pressure corresponding to the assumed planetary radius. To compare our high-resolution (R$\sim$30,000) spectra to the NIRISS/SOSS observations we follow the model binning strategy presented in \citet{Pinhas2018}.

\subsubsection{POSEIDON}
\label{subsubsec:POSEIDON}

The third atmospheric retrieval code we employ is \textsc{POSEIDON} \citep{MacDonald2017,MacDonald2023}. \textsc{POSEIDON} is a well-established atmospheric modelling and spectral retrieval code that was recently released as an open-source\footnote{\textsc{POSEIDON} is available here: \url{https://github.com/MartianColonist/POSEIDON}} Python package \citep{MacDonald2023}. The radiative transfer technique underlying \textsc{POSEIDON}'s transmission spectrum forward model is described in \citet{MacDonald2022}. Our \textsc{POSEIDON} retrieval samples the parameter space using the Bayesian nested sampling algorithm MultiNest, deployed via its Python wrapper \textsc{PyMultiNest} \citep{feroz09,buchner14}.

Our WASP-96\,b \textsc{POSEIDON} retrieval analysis employs a 19-parameter  model accounting for non-isothermal pressure-temperature profiles, inhomogeneous clouds and hazes, and the eight common chemical species described above. One parameter encodes the planetary radius at a 10\,mbar reference radius. The five-parameter PT profile follows the prescription in \citet{Madhusudhan2009}, modified to place the reference temperature parameter at 10\,mbar. The four-parameter inhomogeneous aerosol model follows \citet{MacDonald2017}. Finally, eight parameters specify the constant-in-altitude free abundances of H$_2$O, CO, CO$_2$, CH$_4$, HCN, NH$_3$, Na, and K. The model constructs an atmosphere ranging from $10^{-8}$--$100$\,bar, with 100 layers uniformly distributed in log-pressure, and assumes a H$_2$ + He-dominated background atmosphere with He/H$_2$ = 0.17. The Bayesian retrieval of this 19-parameter space used 1,000 \textsc{PyMultiNest} live points. 


At each location in the parameter space, \textsc{POSEIDON} computed WASP-96\,b transmission spectra at a resolution of $R =$ 20,000 from 0.55--2.9\,µm. The radiative transfer uses opacity sampling of high-resolution pre-computed cross sections ($R \sim 10^6$) from the following line list sources: H$_2$O \citep{Polyansky2018}, CO \citep{Li2015}, CO$_2$ \citep{Tashkun2011}, CH$_4$ \citep{Yurchenko2017}, HCN \citep{Barber2014}, NH$_3$ \citep{Coles2019}, Na \citep{Ryabchikova2015}, and K \citep{Ryabchikova2015}. We additionally include continuum opacity from H$_2$ and He CIA \citep{Karman2019} and H$_2$ Rayleigh scattering \citep{Hohm1994}. We convolve each $R =$ 20,000 model spectrum with the instrument Point Spread Function (PSF), before binning down to the resolution of the observations (here, $R = 125$) to compute the likelihood of each parameter combination. We treat NIRISS/SOSS orders 1 and 2 separately during the convolution and binning procedure, accounting for their different intrinsic PSFs and instrument transmission functions.

\subsubsection{PyratBay}

Lastly, we also employed \textsc{PyratBay}, the Python Radiative-transfer in a Bayesian framework. \textsc{PyratBay}\footnote{The open-source \textsc{PyratBay} code can be found here: \url{https://pyratbay.readthedocs.io/en/latest/}} is an open-source framework for exoplanet atmospheric modelling, spectral synthesis, and Bayesian retrieval. It utilizes the most up to date line-by-line opacity sources from ExoMol \citep{TennysonEtal2016jmsExomol}, HITEMP \citep{Rothman+10}, atomic species Na, K \citep{BurrowsEtal2000} and collision-induced opacities of H$_2$-H$_2$ \citep{Borysow2001, Borysow2002} and H$_2$-He pairs \citep{BorysowEtal1988, BorysowEtal1989, BorysowFrommhold1989}. For effective use in retrieval, we compress these large databases (while retaining information from the dominating line transitions), using the available package \citep{Cubillos2017apjCompress}. 
To model the vertical temperature structure, we implement three parameterization schemes: isothermal, and \citet{Line2013} and \citet{Madhusudhan2009} prescriptions. This retrieval framework also implements a self-consistent 1D radiative-convective equilibrium scheme \citep{MalikEtal2017ajHELIOS}, the classic ``power law+gray'' prescription, a
``single-particle-size'' haze profile, a ``patchy cloud'' prescription for transmission geometry \citep{Line2016}, and two
complex Mie-scattering cloud models. The first is a fully self-consistent microphysical kinetic cloud model of \citet{Helling2006} which follows the formation of seed particles, growth of various solid materials, evaporation, gravitational settling,
elemental depletion and replenishment \citep{BlecicEtal2020-DRIFT}. The other is a parametrized Mie-scattering thermal stability cloud model \citep{KilpatrickEtal2018apjWASP63bWFC3, VenotEtal2020-JWST-WASP-43b}. 

In this work, we assumed the atmosphere of WASP-96\,b to be hydrogen dominated (He/H$_2$ = 0.17) and include collision-induced absorption of H$_2$-H$_2$  and H$_2$-He. We included molecular opacity sources of H$_2$O \citep{Polyansky2018}, CH$_4$ \citep{HargreavesEtal2020}, NH$_3$ \citep{YurchenkoEtal2011, YurchenkoEtal2015}, HCN \citep{HarrisEtal2006, HarrisEtal2008}, CO \citep{Li2015}, and CO$_2$ \citep{Rothman2010} and resonant-line cross-sections of Na and K. In addition, we account for the Rayleigh-scattering cross-section of H$_2$ \citep{DalgarnoWilliams1962} and an unknown haze particulate, by applying a power law prescription of \citet{Lecavelier2008}. Our radiative transfer routine uses opacity sampling of high-resolution pre-computed cross sections tables generated at a resolution of R$\sim$4$\times$10$^7$, calculates the transmission spectra at $R =$ 20,000, and computes the likelihood of each model by binning it down to resolution of $R = 125$. We generated the atmosphere between $10^{-9}$--$100$\,bar, with 81 layers uniformly distributed in log-pressure, retrieving in addition to the constant-with-altitude molecular and alkali volume mixing ratios listed above, also the \citet{Lecavelier2008} haze parameters and the planetary radius at the reference pressure of 0.1\,bar. To find the best modelling setup we tested our available temperature parametrizations and the full range of the cloud models from simple to complex Mie-scattering clouds, assuming species expected to be seen on this temperature regimes. We compared these models using the Bayesian Information criteria \citep[BIC][]{Liddle2007}. We found the lowest BIC for the model assuming \citet{Madhusudhan2009} temperature prescription with patchy opaque cloud deck and hazes, accounting for both \citet{Lecavelier2008} and \cite{DalgarnoWilliams1962} haze particles and opacities from H$_2$O, CO$_2$, CO, Na and K. To explore the phase space of these parameters, we have coupled our atmospheric model with the Bayesian Nested Sampling algorithm \textsc{PyMultiNest} \citep{feroz09,buchner14} and the Multi-core Markov-chain Monte Carlo code \textsc{MC3} \citep{CubillosEtal2016}. Both algorithms returned the same constraints.

\section{Results}
\label{sec: Results}

In this section we present the results from our retrieval analysis. We also discuss the impact the resolution of the data has on our inferred abundances.

\subsection{Retrievals}
\label{sec: Retrievals}

Using the frameworks described above, we infer the atmospheric properties of WASP-96\,b using the NIRISS/SOSS observations binned to four different constant resolutions (R = 125, 250, 500, and pixel level). As discussed below (see Section~\ref{Resolution}), we find our inferences robust regardless of the resolution of the binned observations. Therefore, we present our results using the R = 125 binned observations for clarity. Our first consideration is the possible presence of clouds and hazes. As described above, our atmospheric frameworks compute scenarios representative of cloud-free atmospheres, hazy atmospheres, cloudy atmospheres, and atmospheres with inhomogeneous cloud and haze cover. Comparing these atmospheric scenarios using their Bayesian evidence and comparing them to a `sigma' scale \citep[e.g.,][]{Benneke2013,Welbanks2021}, we find a 6$\sigma$ model preference for inhomogeneous clouds and hazes over simple cloud-free atmospheres. However, we note that it is primarily a Rayleigh scattering slope which we detect as opposed to any opacity from a grey cloud deck (see Section~\ref{sec: clouds params}). We thus limit our discussion to the ``inhomogeneous clouds and hazes'' model runs moving forwards. 

The use of more complex prescriptions separating the spectroscopic effects of clouds from those of hazes across inhomogeneous terminators \citep[e.g.,][]{Welbanks2021}, may result in lower model preferences but consistent inferred atmospheric properties. The full retrieved posterior distributions for Aurora, POSEIDON, PyratBay and CHIMERA can be found in Figures \ref{fig:aurora_corner}, \ref{fig:poseidon_corner}, \ref{fig:pyratbay_corner} and \ref{fig:chimera_corner} respectively.


\subsubsection{Retrieved Abundances}

\begin{figure*}
    \centering
    \includegraphics[width=1.\textwidth]{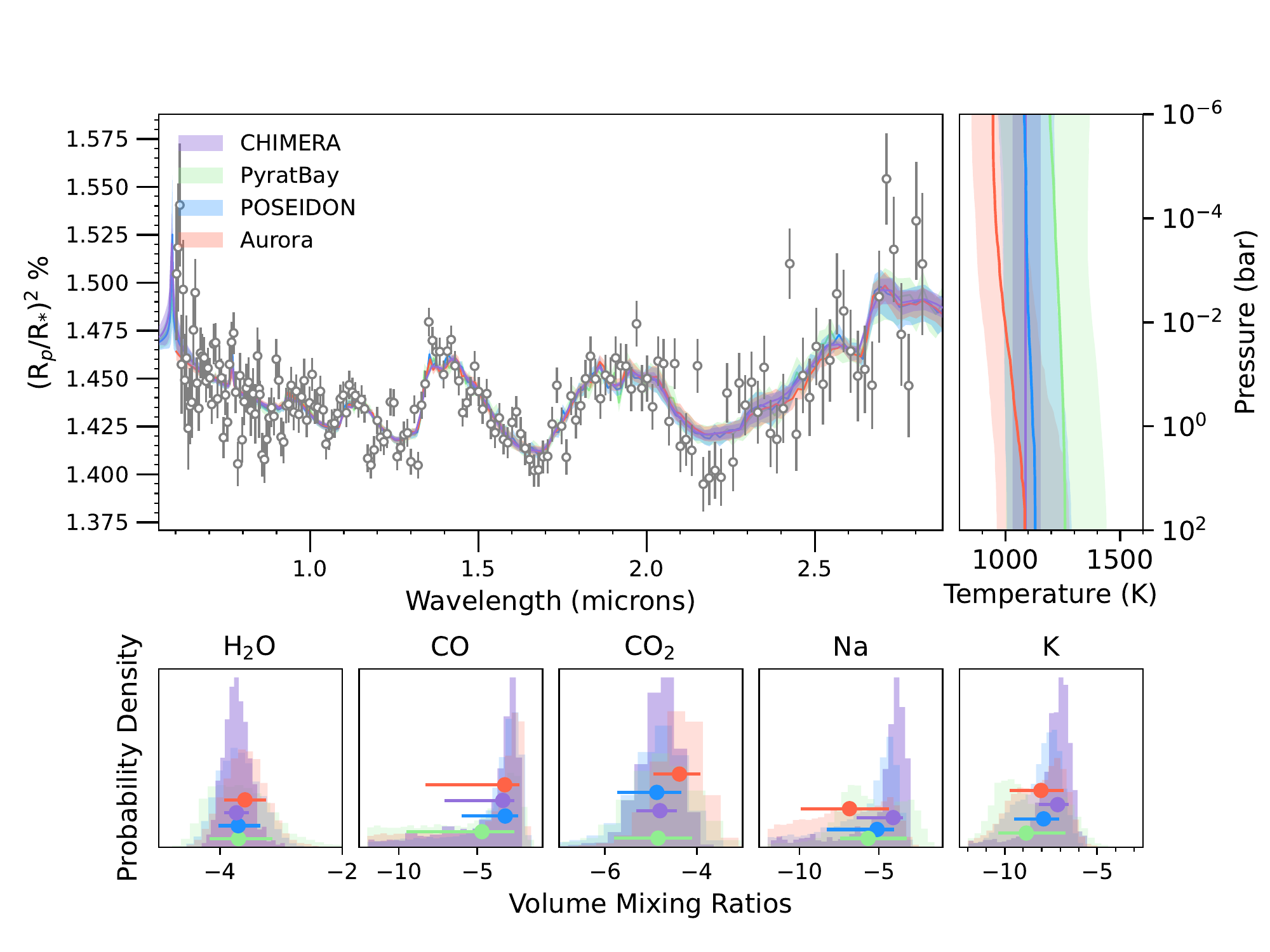}
    \caption{\emph{Top}: Best fit model and best fit temperature profiles, both with 1$\sigma$ error envelope. The models have the following colours: CHIMERA = purple, Pyratbay = green, POSEIDON = blue and Aurora = red. The data is binned to R=125. 
    \emph{Bottom}: Posterior distribution of each molecule that had some constraint, with the same colour coordination as the best fit models. The horizontal line indicates the 1$\sigma$ range. The full 2D corner plots are presented in Appendix~\ref{sec: posteriors}.}
    \label{fig:WASP-96b_Results}
\end{figure*}

The results for the inhomogeneous haze and cloud model runs for all four frameworks are presented in Figure~\ref{fig:WASP-96b_Results}, where we present the best-fit transmission spectra, thermal structure and posteriors for H$_2$O, CO, CO$_2$, Na and K. We do not present the posteriors for NH$_3$, HCN or CH$_4$ as they remain mostly unconstrained given existing observations. The retrieved abundances from all codes are summarized in Table~\ref{tab:results_table}, and generally remain consistent within 1-$\sigma$, demonstrating that the retrieved atmospheric properties are robust against different model implementations. 
They are also largely consistent with a solar metallicity atmosphere, in agreement with the interpretation of  Radica et al.~(submitted) using self-consistent radiative thermochemical equilibrium models.  

\begin{figure*}
    \centering
    \includegraphics[width=0.49\textwidth]{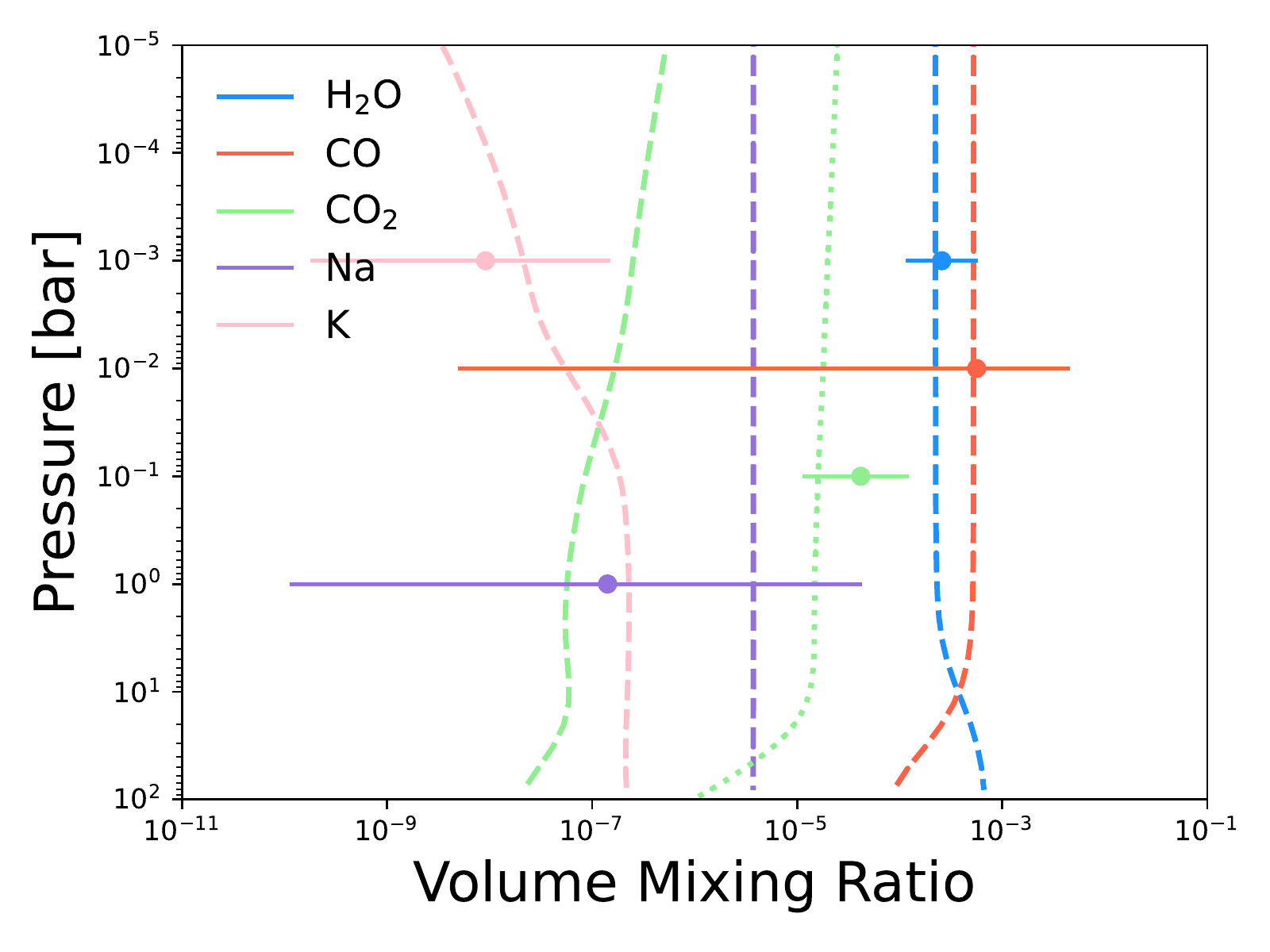}
    \includegraphics[width=0.49\textwidth]{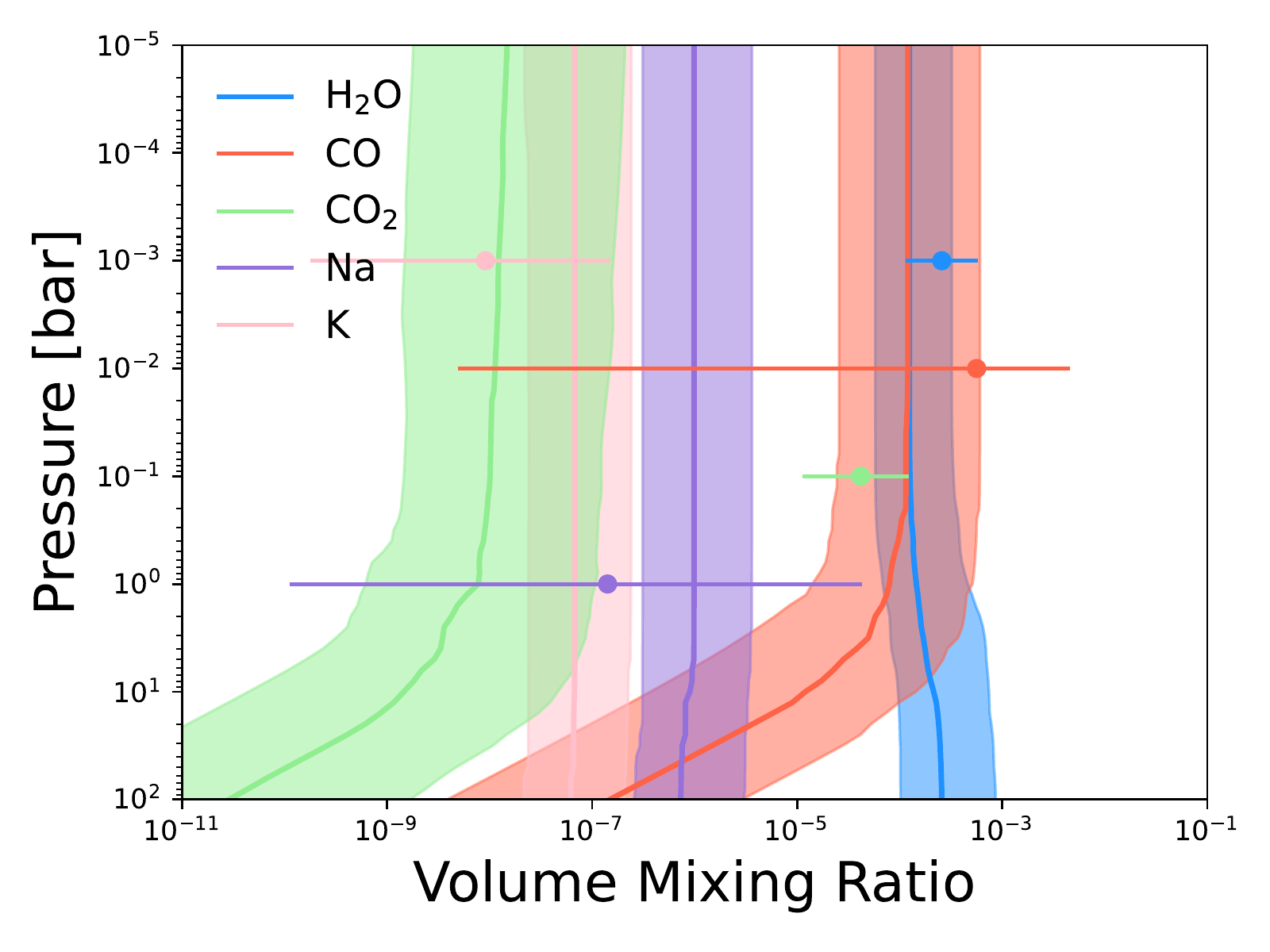}
    \caption{Left: Dashed lines show the Vertical volume mixing ratios obtained from the best-fit ScCHIMERA model in Radica et al.,~(submitted) compared to the horizontal lines which represent the retrieved abundances from the Aurora framework. The dotted line shows 10x solar metalicity. Right: The best-fit retrieved vertical volume mixing ratios obtained from the chemical equilibrium retrieval, the shading representing the 1$\sigma$ uncertainty. These are compared to retrieved results from the free retrieval from Aurora. We note that the vertical location of the retrieved free abundances are arbitrary and do not represent the region probed.}
    \label{fig:chemical_profiles}
\end{figure*}
\begin{table*}
\begin{tabular}{@{}llllll@{}}
\toprule
 &
  log(H$_2$O) &
  log(CO) &
  log(CO$_2$) &
  log(Na) &
  log(K) \\ \midrule
\multicolumn{1}{|l|}{Aurora} &
  \multicolumn{1}{l|}{$-3.59^{+ 0.35 }_{- 0.35 }$} &
  \multicolumn{1}{l|}{$-3.25^{+ 0.91 }_{- 5.06 }$} &
  \multicolumn{1}{l|}{$-4.38^{+ 0.47 }_{- 0.57 }$} &
  \multicolumn{1}{l|}{$-6.85^{+ 2.48 }_{- 3.10 }$} &
  \multicolumn{1}{l|}{$-8.04^{+ 1.22 }_{- 1.71 }$} \\ 
\multicolumn{1}{|l|}{CHIMERA} &
  \multicolumn{1}{l|}{$-3.73^{+ 0.21 }_{- 0.20 }$} &
  \multicolumn{1}{l|}{$-3.39^{+ 0.74 }_{- 3.71 }$} &
  \multicolumn{1}{l|}{$-4.80^{+ 0.37 }_{- 0.52 }$} &
  \multicolumn{1}{l|}{$-4.10^{+ 0.60 }_{-2.31 }$} &
  \multicolumn{1}{l|}{$-7.14^{+0.60}_{-1.02}$} \\ 
\multicolumn{1}{|l|}{POSEIDON} &
  \multicolumn{1}{l|}{$-3.70^{+0.36}_{-0.32}$} &
  \multicolumn{1}{l|}{$-3.22^{+0.81}_{-2.83}$} &
  \multicolumn{1}{l|}{$-4.87^{+ 0.54}_{-0.86}$} &
  \multicolumn{1}{l|}{$-5.13^{+1.07}_{-3.13}$} &
  \multicolumn{1}{l|}{$-7.90^{+0.85}_{-1.59}$} \\ 
\multicolumn{1}{|l|}{PyratBay} &
  \multicolumn{1}{l|}{$-3.70^{+0.56}_{-0.48}$} &
  \multicolumn{1}{l|}{$-4.7^{+2.1}_{-4.8}$} &
  \multicolumn{1}{l|}{$-4.84^{+ 0.75}_{-0.96}$} &
  \multicolumn{1}{l|}{$-5.7^{+2.5}_{-1.8}$} &
  \multicolumn{1}{l|}{$-8.8^{+2.1}_{-1.5}$} \\ 
\multicolumn{1}{|l|}{Solar (1200K @ 1mbar)} &
  \multicolumn{1}{l|}{$-3.37$} &
  \multicolumn{1}{l|}{$-3.27$} &
  \multicolumn{1}{l|}{$-6.71$} &
  \multicolumn{1}{l|}{$-5.42$} &
  \multicolumn{1}{l|}{$-6.61$} \\ 
\end{tabular}
\caption{Retrieved abundances and their accompanying error for our baseline model. We present the results from all of the retrieval frameworks. In the final row we present the abundance of each species calculated at solar metalicity and C/O. The elemental abundances were obtained from \citet{Lodders2002}. We present the vertical volume mixing ratio for these species and compare them to the retrieved values in Figure \ref{fig:chemical_profiles}.}
\label{tab:results_table}
\end{table*}

Using the Aurora framework, we then assess the detection significance of each molecule. This is done by computing the Bayesian evidence for a model without each molecule and comparing to the original model with all species included. We present the breakdown in Table \ref{tab:results_sigma}.

\begin{table}
\begin{tabular}{c|cc}
\hline
Chemical Species                 & log(VMR)                          & Detection significance ($\sigma$) \\ \hline
H$_2$O      & -3.59 $^{+ 0.35 }_{- 0.35 }$ & 16.8                              \\
CO          & -3.25 $^{+ 0.91 }_{- 5.06 }$ & 1.72                              \\
CO$_2$      & -4.38 $^{+ 0.47 }_{- 0.57 }$ & 2.88                              \\
Na         & -6.85$^{+ 2.48 }_{- 3.10 }$  & 1.24                              \\
K           & -8.04$^{+ 1.22 }_{- 1.71 }$  & 2.02                              \\
\hline
Clouds and Hazes & -                            & 6.69                             
\end{tabular}
\caption{Retrieved abundances from Aurora and their accompanying detection significance. We also present the detection significance of using our cloud+haze model compared to a cloud free model.}
\label{tab:results_sigma}
\end{table}

As a final test, we perform a chemically consistent retrieval on the same data using CHIMERA in order to directly retrieve the atmosphere log(C/O) and log(Met). Like the free retrieval, we fit for our Simple Haze + Cloud model. We find the log(C/O) = $-0.30^{+0.17}_{-0.37}$ and log(Met) = $-0.63^{+0.64}_{-0.44}$, where solar values are log(C/O) = $-0.26$ and log(Met) = 0. We present the full posterior distribution of this simulation in Figure \ref{fig:chimera_corner_CE}. Therefore we find that the data is consistent with a model that has a solar C/O ratio within 1$\sigma$ and a solar metallicity within 1$\sigma$. These results are consistent with the modelling work presented in Radica et al.~(submitted). We further demonstrate the consistency with Radica et al.,~(submitted) in Figure \ref{fig:chemical_profiles}, the left panel compares the free retrieved results compared to the volume mixing ratios obtained from the best fitting model in Radica et al.,~(submitted), it can be seen that the abundances obtained in our free retrieval are consistent with these profiles. The outlier is the abundance of CO$_2$, which we find to be consistent with a volume mixing ratio of 10x solar. The right panel shows the retrieved volume mixing ratios for the chemical equilibrium framework, these are again consistent with the free retrieval and the models of Radica et al.,~(submitted).

\subsubsection{Retrieved Cloud Parameters}
\label{sec: clouds params}

We describe in more detail the model preference for inhomogeneous clouds and hazes over the cloud free model described above. The models considering the presence of inhomogeneous clouds and hazes suggest a large fraction ($\gtrsim$70\% i.e., $\theta=0.88 ^{+0.09}_{-0.18}$ Aurora; 0.74$^{+0.08}_{-0.08}$ CHIMERA; 0.91$^{+0.07}_{-0.17}$ POSEIDON; 0.81$^{+0.15}_{-0.15}$ PyratBay) of the planetary terminator covered by either clouds or scattering hazes. However, the retrieved pressure at which the cloud deck is present is consistently high ($\log_{10}(\text{P}_\text{cloud})=0.39 ^{+ 1.04 }_{- 1.08}$ Aurora;  0.38$^{+1.05}_{-1.09}$ POSEIDON; 0.2$^{+1.2}_{-1.2}$ PyratBay) suggesting that the spectroscopic impact of these gray clouds is minimal. Similarly, the low cloud opacity (e.g., log($\kappa_\text{cloud}$) = -32.66$^{+1.62}_{-1.48}$) retrieved by our CHIMERA analysis suggests low impact due to clouds. 

On the other hand, our inferred haze scattering properties suggest they make a significant contribution in our WASP-96\,b observations. While the scattering slope is retrieved to be largely Rayleigh-like (i.e.t, $\gamma=-4.00 ^{+ 0.76 }_{- 1.01 }$ Aurora; $-4.31^{+0.80}_{-0.22}$ CHIMERA;  $-3.75^{+0.68}_{-0.92}$ POSEIDON; $-4.5^{+1.1}_{-1.4}$ PyratBay), the slope is enhanced by more than one order of magnitude ($\log_{10}(\alpha)= 1.85 ^{+0.73}_{-0.47}$ Aurora; 1.70$^{+0.60}_{-0.41}$ POSEIDON; 2.49$^{+0.95}_{-0.77}$ PyratBay). The inferences from the chemical equilibrium retrievals with CHIMERA remain largely in agreement and suggestive of spectroscopic signatures of Rayleigh scattering rather than clouds (e.g., log$_{10}$($\kappa_\text{cloud}$) = $-33.21^{+1.20}_{-1.11}$, $\gamma$ = $-3.31^{+0.43}_{-0.50}$, log$_{10}$($\alpha$) = $-1.39^{+0.26}_{-0.27}$ and f = 0.93$^{+0.05}_{-0.10}$).
All models thus tell the story of an atmosphere with small aerosol particles that produce a Rayleigh scattering slope at short wavelengths, but no evidence for a grey cloud deck, which, as is also the case with our chemical inferences above, is consistent with the interpretation of Radica et al.~(submitted).

\subsection{Resolution Testing}
\label{Resolution}

Atmospheric retrievals can be computationally demanding, and the spectral resolution of the forward model is a large factor in determining the speed of the calculation. To thoroughly study the spectrum of an exoplanet atmosphere, one needs to perform multiple retrieval studies, each study requiring on the order of 10$^4$ to 10$^5$ model calculations; which can become unfeasible at the native R$\sim$700 resolution of NIRISS/SOSS. In this section, we seek to answer the question: Do we infer the same abundances if we bin the native resolution data to lower resolutions?

\begin{figure*}
    \centering
    \includegraphics[width=0.7\textwidth]{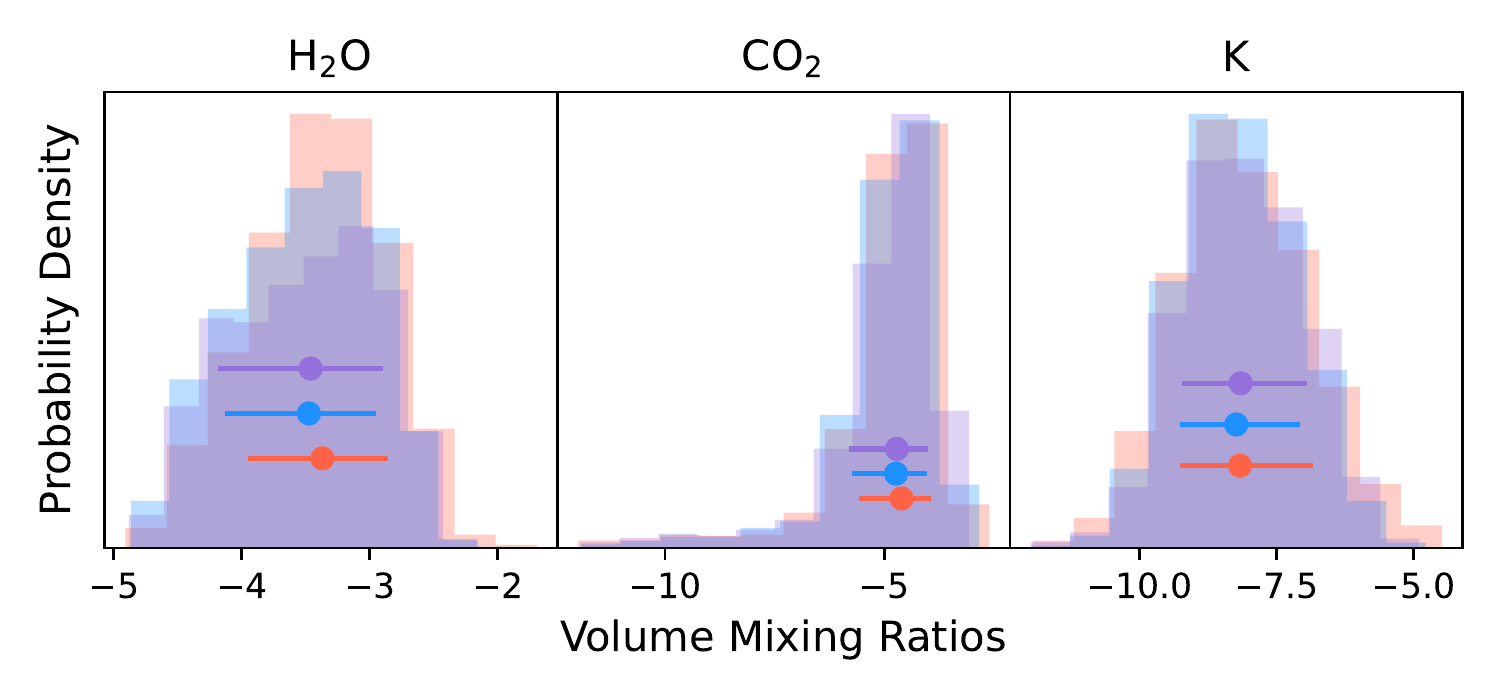}
    \caption{Retrieved posterior distributions on the chemical composition of WASP-96\,b's atmosphere from our resolution test. We binned our observations to R=500, R=250, and R=125 to explore if this binning down of the data causes a loss of information. Posteriors for R=125 are shown in red, blue is R=250 and purple is R=500. The points and error bars show the median retrieved value, and 1$\sigma$ credible interval for each test. A retrieval on each resolution yields consistent abundances to well within 1$\sigma$, allowing us to conclude that no information is lost when binning our WASP-96\,b NIRISS/SOSS transmission observations.}
    \label{fig:WASP-96b_Resolution}
\end{figure*}

To answer this we perform a retrieval analysis on three different transmission spectrum resolutions: R = 125, R = 250, and R = 500, shown in Figure~\ref{fig:binned_spectra}. We use the same parameterised model presented in Figure \ref{fig:WASP-96b_Results} and correlated-$k$ tables calculated at R=3000; hence the model has a resolution six times greater than the maximum data resolution. We find that the retrieved abundances for data with a resolution of R=125 are the same as with a resolution of R=500. Hence, no information is lost when binning the data. We present the posteriors of H$_2$O, CO$_2$, and K in Figure~\ref{fig:WASP-96b_Resolution}. The colours correspond to those in Figure~\ref{fig:binned_spectra}.

\section{Discussion}
\label{sec: Discussion}

Since the first ground-based observations of WASP-96\,b by \citet{Nikolov2018} revealed pressure-broadened Na wings, the planet has held the unique privilege of being one of the few "cloud-free" exoplanets known. Subsequent studies \citep{Nikolov2022, McGruder2022} added HST/WFC3 transit depths, as well as additional ground-based transmission observations from Magellan/IMACS, however the conclusion of the cloud-free nature of WASP-96\,b's upper atmosphere remained unchanged. The GCM models of \citet{Samra2022}, though, found that the terminator region of WASP-96\,b should be entirely covered in clouds given the temperature structure of the planet. Moreover, they show that cloudy transmission spectra can provide an equally good fit to the ensemble of transmission data analyzed in \citet{Nikolov2022}.

Our two independent GCM models also predict that clouds should be able to form at the terminator of WASP-96\,b in the pressure regions probed by transmission spectroscopy (see Figure~\ref{fig:transmission_spectrum_gcm}). These models predict that the atmosphere is likely dominated by MnS and Na$_2$S clouds. MgSiO$_3$ clouds should form in the deep layers of the atmosphere and would be observable only if the vertical mixing was extremely large to easily replenish the upper atmosphere in cloud-forming material, an assumption that is inherent to the \citet{Samra2022} calculation.


{One solution to this discrepancy could be that smaller particles than predicted by \citet{Samra2022} form in larger quantities at low pressures in WASP-96b's atmosphere. These could be composed of $Na_2S$ or KCl which would naturally form at much lower pressures than the silicate clouds that dominate the cloud composition in the 100 to 10 mbar range. However, the detection of sodium and potassium in WASP-96b's atmosphere seems to rule out this possibility. MnS is another candidate for forming clouds at low pressures~\citep{Parmentier2016,Morley2012}, however, \citet{Gao2020} predicts that the nucleation rates for MnS are so low that they should hardly form. Another option would be the formation of a high altitude haze layer formed of photochemically produced particles. Photochemistry is known to naturally form small particles at low pressures that can produce strong scattering slopes~\citep{Lavvas2017,Kawashima2019,Helling2020,Steinrueck2021}. Additional information about the cloud composition could be gathered by targeting the resonant features of the cloud-forming material in the JWST/MIRI LRS bandpass.

We further note that our detection of a strong scattering slope in the optical is partially degenerate with the abundance of gaseous sodium in the atmosphere.  Indeed when a scattering slope is not included in the retrievals, we obtain an unphysical alkali abundances (e.g., $\log_{10}(\text{Na}) = -2.54^{0.28}_{-0.34}$ with CHIMERA). However, including enhanced Rayleigh scattering, the Na abundances drops to slightly super-solar to solar values, in agreement with \citet{Nikolov2022}. Our inferred abundances of Na and of the presence of a scattering slope therefore needs to be carefully interpreted because of this degeneracy, driven by the fact that the NIRISS/SOSS bandpass cuts off at 0.6\,µm, and is therefore only able to probe the red wing of the Na feature. Without fully resolving the Na feature peak, it is difficult to differentiate between a slope caused by a Rayleigh scattering haze, or the red wing of a broadened Na feature. More work needs to be conducted to further understand this degeneracy in the context of observations with NIRISS/SOSS.

\subsection{Comparison to Radica et al., (submitted)}
A suite of forward models were compared to the data in our companion paper (Radica et al., submitted). Three different grids of models were used: PICASO, ATMO and ScCHIMERA, producing a picture of an atmosphere that has a metallicity of 1 -- 5$\times$ solar and a solar C/O. Our free retrieval results demonstrate that we are obtaining an abundance of H$_2$O that is consistent with solar values and a CO$_2$ abundance that is super solar, this demonstrates that our results are consistent with Radica et al., submitted. Similarly to Radica et al., submitted, we need to invoke enhanced Rayleigh scattering slope to match the observations at the shortest wavelengths, but find no spectroscopic impact from a grey cloud deck. We compare the vertical volume mixing ratios obtained from the best fit ScCHIMERA model with our retrieved results in Figure \ref{fig:chemical_profiles} which shows we are obtaining a consistent picture of the atmosphere.

\section{Conclusions}
\label{sec: Conclusions}

In this paper we have performed a detailed atmospheric characterisation of WASP-96\,b using the transmission spectrum obtained with NIRISS/SOSS as part of the Early Release Observations, and first presented in Radica et al.~(submitted). 

We ran GCM simulations in order to model the planet's atmosphere using the SPARC MIT/gcm and the UM. These clear-sky models are able to well fit the spectrum redward of 1.3\,µm, and favour an atmosphere with solar metallicity. However, blueward of 1.3\,µm, the GCMs under predict the observed transit depths, likely indicating missing opacities such as a scattering haze.

We then performed a suite of retrievals using four different modelling frameworks: CHIMERA, Aurora, PyratBay, and POSEIDON. We find that a model with patchy clouds and hazes best describe the data and that each framework produces results which are consistent within 1$\sigma$. We report the retrieved abundances from Aurora as $\log_{10}(\text{H}_2\text{O}) = -3.59 ^{+ 0.35 }_{- 0.35 }$, $\log_{10}(\text{K}) = -8.04 ^{+ 1.22 }_{- 1.71 }$, $\log_{10}(\text{CO}) = -3.25$, and $\log_{10}(\text{CO}_2) = -4.38 ^{+ 0.47 }_{- 0.57 }$. We find a large tail in the posterior CO, so we describe this abundance as an upper limit. Further transmission observations with JWST, particularly with NIRSpec G395H are necessary to more accurately constrain the abundance of CO.

The retrieved abundance of H$_2$O is consistent with \citet{Yip2021} and \citet{McGruder2022}. Our precision is $\sim$10$\times$ better than \citet{McGruder2022} and $\sim$4$\times$ better \citet{Yip2021}. Our range of retrieved abundances of Na is consistent with \citet{Nikolov2022}, \citet{McGruder2022}, \citet{Yip2021} and \citet{Welbanks2019}, however given that NIRISS' wavelength coverage does not capture the complete Na feature, this results in a degeneracy between the abundance of Na and a Rayleigh scattering slope. This is also reflected in the extremely low detection significance for Na (1.24\,$\sigma$). We therefore caution against any strong interpretations of this Na abundance. We also report a constrained abundance of potassium, although with only a marginal detection significance ($\sim$2$\sigma$), in the atmosphere of WASP-96\,b which was not found in previous studies due to the lower resolution of the optical data. The strong potassium constraint in the atmosphere of WASP-39\,b from NIRISS/SOSS \citep{feinsteinw39}, and the tentative detection here, demonstrates how powerful this instrument is to study alkali metals, and opens the door for a new tracer of formation history, the K/O ratio \citep{feinsteinw39}.

Our chemically consistent retrievals favour an atmosphere that has a solar C/O ratio within 1$\sigma$ and solar metallicity within 1$\sigma$. We find the log(C/O) = $-0.30^{+0.17}_{-0.37}$ and log(Met) = $-0.63^{+0.64}_{-0.44}$, where solar values are log(C/O) = $-0.26$ and log(Met) = 0. This is consistent with the GCM models and the grid models in Radica et al.~(submitted) which favour an atmosphere that is 1$\times$ solar C/O and 1 -- 5$\times$ solar M/H.

We explore the appropriate resolution to study observations obtained with NIRISS/SOSS. We find that binning the data from native to R=125 does not impact the inferred abundances. This is useful, given that retrievals at native resolution are computational demanding. In the era of JWST, we need to explore more complex models, which are computational demanding in themselves, therefore we should trade data resolution for model complexity.

Finally, it is critical to note that the previous studies retrieved on a transmission spectrum created through the combination of multiple instruments, with \textit{six transits} required to construct the \citet{Nikolov2022} spectrum. The NIRISS/SOSS transmission spectrum we have presented here was obtained with \textit{one single transit observation}, further highlighting the undeniable potential of JWST to unveil atmospheres of transiting exoplanets.

\section*{Acknowledgements}
We thank the anonymous reviewer for their detailed feedback which greatly improved our manuscript. JT thanks the John Fell Fund and the Canadian Space Agency for financially supporting this work. MR acknowledges financial support from the National Sciences and Research Council of Canada (NSERC), the Fonds de Recherche du Qu\'ebec - Nature et Technologies (FRQNT), and the Institut Trottier de recherche sur les exoplanètes (iREx). LW and RJM acknowledge support for this work provided by NASA through the NASA Hubble Fellowship grant \#HST-HF2-51496.001-A and \#HST-HF2-51513.001, respectively, awarded by the Space Telescope Science Institute, which is operated by the Association of Universities for Research in Astronomy, Inc., for NASA, under contract NAS5-26555.\\
JB acknowledges the support received in part from the NYUAD IT High Performance Computing resources, services, and staff expertise. \\
MZ was supported through a UKRI Future Leaders Fellowship MR/T040866/1. MZ's Met Office \textsc{Unified Model} simulations were performed using the DiRAC Data Intensive service at Leicester, operated by the University of Leicester IT Services, which forms part of the STFC DiRAC HPC Facility (\url{www.dirac.ac.uk}). The equipment was funded by BEIS capital funding via STFC capital grants ST/K000373/1 and ST/R002363/1 and STFC DiRAC Operations grant ST/R001014/1. DiRAC is part of the National e-Infrastructure. A portion of this analysis was carried out on the High Performance Computing resources at New York University Abu Dhabi and resources provided by the Research Computing at Arizona State University.\\
This work is based observations made with the NASA/ESA/CSA James Webb Space Telescope. The data were obtained from the Mikulski Archive for Space Telescopes at the Space Telescope Science Institute, which is operated by the Association of Universities for Research in Astronomy, Inc., under NASA contract NAS 5-03127 for JWST.\\

\section*{Software}
\begin{itemize}
    \renewcommand\labelitemi{--}
    \item \texttt{astropy}; \citet{astropy:2013, astropy:2018}
    \item \texttt{matplotlib}; \citet{Hunter:2007}
    \item \texttt{numpy}; \citet{harris2020array}
    \item \texttt{scipy}; \citet{2020SciPy-NMeth}
    \item Met Office \textsc{Unified Model} materials were produced using Met Office Software.
    \item \texttt{PyMultiNest} \citep{buchner14}
    \item \texttt{corner} \citep{corner}
\end{itemize}

\section*{Data Availability}
All data used in this study is publicly available from the Barbara A. Mikulski Archive for Space Telescopes\footnote{https://mast.stsci.edu/portal/Mashup/Clients/Mast/Portal.html}. The models generated in this paper can be made available on request.

\bibliographystyle{mnras}
\bibliography{bib, software} 

\begin{thebibliography}{}
\makeatletter
\relax
\def\mn@urlcharsother{\let\do\@makeother \do\$\do\&\do\#\do\^\do\_\do\%\do\~}
\def\mn@doi{\begingroup\mn@urlcharsother \@ifnextchar [ {\mn@doi@}
  {\mn@doi@[]}}
\def\mn@doi@[#1]#2{\def\@tempa{#1}\ifx\@tempa\@empty \href
  {http://dx.doi.org/#2} {doi:#2}\else \href {http://dx.doi.org/#2} {#1}\fi
  \endgroup}
\def\mn@eprint#1#2{\mn@eprint@#1:#2::\@nil}
\def\mn@eprint@arXiv#1{\href {http://arxiv.org/abs/#1} {{\tt arXiv:#1}}}
\def\mn@eprint@dblp#1{\href {http://dblp.uni-trier.de/rec/bibtex/#1.xml}
  {dblp:#1}}
\def\mn@eprint@#1:#2:#3:#4\@nil{\def\@tempa {#1}\def\@tempb {#2}\def\@tempc
  {#3}\ifx \@tempc \@empty \let \@tempc \@tempb \let \@tempb \@tempa \fi \ifx
  \@tempb \@empty \def\@tempb {arXiv}\fi \@ifundefined
  {mn@eprint@\@tempb}{\@tempb:\@tempc}{\expandafter \expandafter \csname
  mn@eprint@\@tempb\endcsname \expandafter{\@tempc}}}

\bibitem[\protect\citeauthoryear{{Alderson} et~al.,}{{Alderson}
  et~al.}{2022}]{Alderson2022}
{Alderson} L.,  et~al., 2022, arXiv e-prints, \href
  {https://ui.adsabs.harvard.edu/abs/2022arXiv221110488A} {p. arXiv:2211.10488}

\bibitem[\protect\citeauthoryear{{Allard}, {Spiegelman}  \&
  {Kielkopf}}{{Allard} et~al.}{2016a}]{Allard+16}
{Allard} N.~F.,  {Spiegelman} F.,   {Kielkopf} J.~F.,  2016a, \mn@doi [\aap]
  {10.1051/0004-6361/201628270}, \href
  {https://ui.adsabs.harvard.edu/abs/2016A&A...589A..21A} {589, A21}

\bibitem[\protect\citeauthoryear{{Allard}, {Spiegelman}  \&
  {Kielkopf}}{{Allard} et~al.}{2016b}]{Allard2016}
{Allard} N.~F.,  {Spiegelman} F.,   {Kielkopf} J.~F.,  2016b, \mn@doi [\aap]
  {10.1051/0004-6361/201628270}, \href
  {https://ui.adsabs.harvard.edu/abs/2016A&A...589A..21A} {589, A21}

\bibitem[\protect\citeauthoryear{{Allard}, {Spiegelman}, {Leininger}  \&
  {Molliere}}{{Allard} et~al.}{2019}]{Allard2019}
{Allard} N.~F.,  {Spiegelman} F.,  {Leininger} T.,   {Molliere} P.,  2019,
  \mn@doi [\aap] {10.1051/0004-6361/201935593}, \href
  {https://ui.adsabs.harvard.edu/abs/2019A&A...628A.120A} {628, A120}

\bibitem[\protect\citeauthoryear{{Astropy Collaboration} et~al.,}{{Astropy
  Collaboration} et~al.}{2013}]{astropy:2013}
{Astropy Collaboration} et~al., 2013, \mn@doi [\aap]
  {10.1051/0004-6361/201322068}, \href
  {http://adsabs.harvard.edu/abs/2013A%26A...558A..33A} {558, A33}

\bibitem[\protect\citeauthoryear{{Astropy Collaboration} et~al.,}{{Astropy
  Collaboration} et~al.}{2018}]{astropy:2018}
{Astropy Collaboration} et~al., 2018, \mn@doi [\aj] {10.3847/1538-3881/aabc4f},
  \href {https://ui.adsabs.harvard.edu/abs/2018AJ....156..123A} {156, 123}

\bibitem[\protect\citeauthoryear{{Barber}, {Strange}, {Hill}, {Polyansky},
  {Mellau}, {Yurchenko}  \& {Tennyson}}{{Barber} et~al.}{2014}]{Barber2014}
{Barber} R.~J.,  {Strange} J.~K.,  {Hill} C.,  {Polyansky} O.~L.,  {Mellau}
  G.~C.,  {Yurchenko} S.~N.,   {Tennyson} J.,  2014, \mn@doi [\mnras]
  {10.1093/mnras/stt2011}, \href
  {http://adsabs.harvard.edu/abs/2014MNRAS.437.1828B} {437, 1828}

\bibitem[\protect\citeauthoryear{{Barstow}, {Changeat}, {Garland}, {Line},
  {Rocchetto}  \& {Waldmann}}{{Barstow} et~al.}{2020}]{Barstow2020}
{Barstow} J.~K.,  {Changeat} Q.,  {Garland} R.,  {Line} M.~R.,  {Rocchetto} M.,
    {Waldmann} I.~P.,  2020, \mn@doi [\mnras] {10.1093/mnras/staa548}, \href
  {https://ui.adsabs.harvard.edu/abs/2020MNRAS.493.4884B} {493, 4884}

\bibitem[\protect\citeauthoryear{{Barstow}, {Changeat}, {Chubb}, {Cubillos},
  {Edwards}, {MacDonald}, {Min}  \& {Waldmann}}{{Barstow}
  et~al.}{2022}]{Barstow2022}
{Barstow} J.~K.,  {Changeat} Q.,  {Chubb} K.~L.,  {Cubillos} P.~E.,  {Edwards}
  B.,  {MacDonald} R.~J.,  {Min} M.,   {Waldmann} I.~P.,  2022, \mn@doi
  [Experimental Astronomy] {10.1007/s10686-021-09821-w}, \href
  {https://ui.adsabs.harvard.edu/abs/2022ExA....53..447B} {53, 447}

\bibitem[\protect\citeauthoryear{{Benneke} \& {Seager}}{{Benneke} \&
  {Seager}}{2013}]{Benneke2013}
{Benneke} B.,  {Seager} S.,  2013, \mn@doi [\apj]
  {10.1088/0004-637X/778/2/153}, \href
  {https://ui.adsabs.harvard.edu/abs/2013ApJ...778..153B} {778, 153}

\bibitem[\protect\citeauthoryear{{Blecic}, {Helling}, {Woitke}, {Dobbs-Dixon}
  \& {Greene}}{{Blecic} et~al.}{2023}]{BlecicEtal2020-DRIFT}
{Blecic} J.,  {Helling} C.,  {Woitke} P.,  {Dobbs-Dixon} S.,   {Greene} T.,
  2023, in prep

\bibitem[\protect\citeauthoryear{{Borysow}}{{Borysow}}{2002}]{Borysow2002}
{Borysow} A.,  2002, \mn@doi [\aap] {10.1051/0004-6361:20020555}, \href
  {https://ui.adsabs.harvard.edu/abs/2002A&A...390..779B} {390, 779}

\bibitem[\protect\citeauthoryear{{Borysow} \& {Frommhold}}{{Borysow} \&
  {Frommhold}}{1989}]{BorysowFrommhold1989}
{Borysow} A.,  {Frommhold} L.,  1989, \mn@doi [\apj] {10.1086/167515}, \href
  {https://ui.adsabs.harvard.edu/abs/1989ApJ...341..549B} {341, 549}

\bibitem[\protect\citeauthoryear{{Borysow}, {Frommhold}  \&
  {Birnbaum}}{{Borysow} et~al.}{1988}]{BorysowEtal1988}
{Borysow} J.,  {Frommhold} L.,   {Birnbaum} G.,  1988, \mn@doi [\apj]
  {10.1086/166112}, \href
  {https://ui.adsabs.harvard.edu/abs/1988ApJ...326..509B} {326, 509}

\bibitem[\protect\citeauthoryear{{Borysow}, {Frommhold}  \&
  {Moraldi}}{{Borysow} et~al.}{1989}]{BorysowEtal1989}
{Borysow} A.,  {Frommhold} L.,   {Moraldi} M.,  1989, \mn@doi [\apj]
  {10.1086/167027}, \href
  {https://ui.adsabs.harvard.edu/abs/1989ApJ...336..495B} {336, 495}

\bibitem[\protect\citeauthoryear{{Borysow}, {Jorgensen}  \& {Fu}}{{Borysow}
  et~al.}{2001}]{Borysow2001}
{Borysow} A.,  {Jorgensen} U.~G.,   {Fu} Y.,  2001, \mn@doi [\jqsrt]
  {10.1016/S0022-4073(00)00023-6}, \href
  {https://ui.adsabs.harvard.edu/abs/2001JQSRT..68..235B} {68, 235}

\bibitem[\protect\citeauthoryear{{Buchner} et~al.,}{{Buchner}
  et~al.}{2014}]{buchner14}
{Buchner} J.,  et~al., 2014, \mn@doi [Astronomy and Astrophysics]
  {10.1051/0004-6361/201322971}, \href
  {https://ui.adsabs.harvard.edu/\#abs/2014Astronomy and
  Astrophysics...564A.125B} {564, A125}

\bibitem[\protect\citeauthoryear{{Burrows}, {Marley}  \& {Sharp}}{{Burrows}
  et~al.}{2000}]{BurrowsEtal2000}
{Burrows} A.,  {Marley} M.~S.,   {Sharp} C.~M.,  2000, \mn@doi [\apj]
  {10.1086/308462}, \href
  {https://ui.adsabs.harvard.edu/abs/2000ApJ...531..438B} {531, 438}

\bibitem[\protect\citeauthoryear{{Coles}, {Yurchenko}  \& {Tennyson}}{{Coles}
  et~al.}{2019}]{Coles2019}
{Coles} P.~A.,  {Yurchenko} S.~N.,   {Tennyson} J.,  2019, \mn@doi [\mnras]
  {10.1093/mnras/stz2778}, \href
  {https://ui.adsabs.harvard.edu/abs/2019MNRAS.490.4638C} {490, 4638}

\bibitem[\protect\citeauthoryear{{Coulombe} et~al.,}{{Coulombe}
  et~al.}{2023}]{CoulombeW18b}
{Coulombe} L.-P.,  et~al., 2023, arXiv e-prints, \href
  {https://ui.adsabs.harvard.edu/abs/2023arXiv230108192C} {p. arXiv:2301.08192}

\bibitem[\protect\citeauthoryear{{Cubillos}}{{Cubillos}}{2017}]{Cubillos2017apjCompress}
{Cubillos} P.~E.,  2017, \mn@doi [\apj] {10.3847/1538-4357/aa9228}, \href
  {https://ui.adsabs.harvard.edu/abs/2017ApJ...850...32C} {850, 32}

\bibitem[\protect\citeauthoryear{{Cubillos} \& {Blecic}}{{Cubillos} \&
  {Blecic}}{2021}]{CubillosBlecic2021MNRAS}
{Cubillos} P.~E.,  {Blecic} J.,  2021, \mn@doi [\mnras]
  {10.1093/mnras/stab1405}, \href
  {https://ui.adsabs.harvard.edu/abs/2021MNRAS.505.2675C} {505, 2675}

\bibitem[\protect\citeauthoryear{{Cubillos} et~al.,}{{Cubillos}
  et~al.}{2016}]{CubillosEtal2016}
{Cubillos} P.,  et~al., 2016, {MC3: Multi-core Markov-chain Monte Carlo code},
  Astrophysics Source Code Library, record ascl:1610.013 (\mn@eprint {ascl}
  {1610.013})

\bibitem[\protect\citeauthoryear{{Dalgarno} \& {Williams}}{{Dalgarno} \&
  {Williams}}{1962}]{DalgarnoWilliams1962}
{Dalgarno} A.,  {Williams} D.~A.,  1962, \mn@doi [\apj] {10.1086/147428}, \href
  {https://ui.adsabs.harvard.edu/abs/1962ApJ...136..690D} {136, 690}

\bibitem[\protect\citeauthoryear{{Darveau-Bernier} et~al.,}{{Darveau-Bernier}
  et~al.}{2022}]{darveau-bernier_atoca}
{Darveau-Bernier} A.,  et~al., 2022, \mn@doi [\pasp]
  {10.1088/1538-3873/ac8a77}, \href
  {https://ui.adsabs.harvard.edu/abs/2022PASP..134i4502D} {134, 094502}

\bibitem[\protect\citeauthoryear{Drummond, Tremblin, Baraffe, Amundsen, Mayne,
  Venot  \& Goyal}{Drummond et~al.}{2016}]{Drummond2016}
Drummond B.,  Tremblin P.,  Baraffe I.,  Amundsen D.~S.,  Mayne N.~J.,  Venot
  O.,   Goyal J.,  2016, \mn@doi [Astronomy and Astrophysics]
  {10.1051/0004-6361/201628799}, 594

\bibitem[\protect\citeauthoryear{Drummond et~al.,}{Drummond
  et~al.}{2020}]{Drummond2020}
Drummond B.,  et~al., 2020, \mn@doi [Astronomy and Astrophysics]
  {10.1051/0004-6361/201937153}, 636

\bibitem[\protect\citeauthoryear{{Feinstein} et~al.,}{{Feinstein}
  et~al.}{2023}]{feinsteinw39}
{Feinstein} A.~D.,  et~al., 2023, \mn@doi [\nat] {10.1038/s41586-022-05674-1},
  \href {https://ui.adsabs.harvard.edu/abs/2023Natur.614..670F} {614, 670}

\bibitem[\protect\citeauthoryear{{Feroz}, {Hobson}  \& {Bridges}}{{Feroz}
  et~al.}{2009}]{feroz09}
{Feroz} F.,  {Hobson} M.~P.,   {Bridges} M.,  2009, \mn@doi [\mnras]
  {10.1111/j.1365-2966.2009.14548.x}, \href
  {http://adsabs.harvard.edu/abs/2009MNRAS.398.1601F} {398, 1601}

\bibitem[\protect\citeauthoryear{Foreman-Mackey}{Foreman-Mackey}{2016}]{corner}
Foreman-Mackey D.,  2016, \mn@doi [The Journal of Open Source Software]
  {10.21105/joss.00024}, 1, 24

\bibitem[\protect\citeauthoryear{{Fortney}}{{Fortney}}{2005}]{Fortney2005}
{Fortney} J.~J.,  2005, \mn@doi [\mnras] {10.1111/j.1365-2966.2005.09587.x},
  \href {https://ui.adsabs.harvard.edu/abs/2005MNRAS.364..649F} {364, 649}

\bibitem[\protect\citeauthoryear{{Freedman}, {Lustig-Yaeger}, {Fortney},
  {Lupu}, {Marley}  \& {Lodders}}{{Freedman} et~al.}{2014}]{Freedman2014}
{Freedman} R.~S.,  {Lustig-Yaeger} J.,  {Fortney} J.~J.,  {Lupu} R.~E.,
  {Marley} M.~S.,   {Lodders} K.,  2014, \mn@doi [\apjs]
  {10.1088/0067-0049/214/2/25}, \href
  {https://ui.adsabs.harvard.edu/abs/2014ApJS..214...25F} {214, 25}

\bibitem[\protect\citeauthoryear{{Fu} et~al.,}{{Fu} et~al.}{2022}]{Fu2022}
{Fu} G.,  et~al., 2022, arXiv e-prints, \href
  {https://ui.adsabs.harvard.edu/abs/2022arXiv221113761F} {p. arXiv:2211.13761}

\bibitem[\protect\citeauthoryear{{Gandhi} \& {Madhusudhan}}{{Gandhi} \&
  {Madhusudhan}}{2017}]{Gandhi2017}
{Gandhi} S.,  {Madhusudhan} N.,  2017, \mn@doi [\mnras]
  {10.1093/mnras/stx1601}, \href
  {http://adsabs.harvard.edu/abs/2017MNRAS.472.2334G} {472, 2334}

\bibitem[\protect\citeauthoryear{{Gandhi} \& {Madhusudhan}}{{Gandhi} \&
  {Madhusudhan}}{2018}]{Gandhi2018}
{Gandhi} S.,  {Madhusudhan} N.,  2018, \mn@doi [\mnras]
  {10.1093/mnras/stx2748}, \href
  {https://ui.adsabs.harvard.edu/abs/2018MNRAS.474..271G} {474, 271}

\bibitem[\protect\citeauthoryear{{Gandhi} et~al.,}{{Gandhi}
  et~al.}{2020}]{Gandhi2020}
{Gandhi} S.,  et~al., 2020, \mn@doi [\mnras] {10.1093/mnras/staa981}, \href
  {https://ui.adsabs.harvard.edu/abs/2020MNRAS.495..224G} {495, 224}

\bibitem[\protect\citeauthoryear{Gao et~al.,}{Gao et~al.}{2020}]{Gao2020}
Gao P.,  et~al., 2020, \mn@doi [Nature Astronomy] {10.1038/s41550-020-1114-3}

\bibitem[\protect\citeauthoryear{{Gharib-Nezhad}, {Iyer}, {Line}, {Freedman},
  {Marley}  \& {Batalha}}{{Gharib-Nezhad} et~al.}{2021}]{Gharib2021}
{Gharib-Nezhad} E.,  {Iyer} A.~R.,  {Line} M.~R.,  {Freedman} R.~S.,  {Marley}
  M.~S.,   {Batalha} N.~E.,  2021, \mn@doi [\apjs] {10.3847/1538-4365/abf504},
  \href {https://ui.adsabs.harvard.edu/abs/2021ApJS..254...34G} {254, 34}

\bibitem[\protect\citeauthoryear{Gordon \& McBride}{Gordon \&
  McBride}{1994}]{gordon1994computer}
Gordon S.,  McBride B.~J.,  1994, Technical report, Computer program for
  calculation of complex chemical equilibrium compositions and applications.
  Part 1: Analysis

\bibitem[\protect\citeauthoryear{{Grimm} et~al.,}{{Grimm}
  et~al.}{2021}]{Grimm2021}
{Grimm} S.~L.,  et~al., 2021, \mn@doi [\apjs] {10.3847/1538-4365/abd773}, \href
  {https://ui.adsabs.harvard.edu/abs/2021ApJS..253...30G} {253, 30}

\bibitem[\protect\citeauthoryear{{Hargreaves}, {Gordon}, {Rey}, {Nikitin},
  {Tyuterev}, {Kochanov}  \& {Rothman}}{{Hargreaves}
  et~al.}{2020}]{HargreavesEtal2020}
{Hargreaves} R.~J.,  {Gordon} I.~E.,  {Rey} M.,  {Nikitin} A.~V.,  {Tyuterev}
  V.~G.,  {Kochanov} R.~V.,   {Rothman} L.~S.,  2020, \mn@doi [\apjs]
  {10.3847/1538-4365/ab7a1a}, \href
  {https://ui.adsabs.harvard.edu/abs/2020ApJS..247...55H} {247, 55}

\bibitem[\protect\citeauthoryear{{Harris}, {Tennyson}, {Kaminsky}, {Pavlenko}
  \& {Jones}}{{Harris} et~al.}{2006}]{HarrisEtal2006}
{Harris} G.~J.,  {Tennyson} J.,  {Kaminsky} B.~M.,  {Pavlenko} Y.~V.,   {Jones}
  H.~R.~A.,  2006, \mn@doi [\mnras] {10.1111/j.1365-2966.2005.09960.x}, \href
  {https://ui.adsabs.harvard.edu/abs/2006MNRAS.367..400H} {367, 400}

\bibitem[\protect\citeauthoryear{{Harris}, {Larner}, {Tennyson}, {Kaminsky},
  {Pavlenko}  \& {Jones}}{{Harris} et~al.}{2008}]{HarrisEtal2008}
{Harris} G.~J.,  {Larner} F.~C.,  {Tennyson} J.,  {Kaminsky} B.~M.,  {Pavlenko}
  Y.~V.,   {Jones} H.~R.~A.,  2008, \mn@doi [\mnras]
  {10.1111/j.1365-2966.2008.13642.x}, \href
  {https://ui.adsabs.harvard.edu/abs/2008MNRAS.390..143H} {390, 143}

\bibitem[\protect\citeauthoryear{Harris et~al.,}{Harris
  et~al.}{2020}]{harris2020array}
Harris C.~R.,  et~al., 2020, \mn@doi [Nature] {10.1038/s41586-020-2649-2}, 585,
  357

\bibitem[\protect\citeauthoryear{{Hartman} et~al.,}{{Hartman}
  et~al.}{2011}]{Hartman2011}
{Hartman} J.~D.,  et~al., 2011, \mn@doi [\apj] {10.1088/0004-637X/726/1/52},
  \href {https://ui.adsabs.harvard.edu/abs/2011ApJ...726...52H} {726, 52}

\bibitem[\protect\citeauthoryear{{Hellier} et~al.,}{{Hellier}
  et~al.}{2014}]{Hellier2014}
{Hellier} C.,  et~al., 2014, \mn@doi [\mnras] {10.1093/mnras/stu410}, \href
  {https://ui.adsabs.harvard.edu/abs/2014MNRAS.440.1982H} {440, 1982}

\bibitem[\protect\citeauthoryear{{Helling} \& {Woitke}}{{Helling} \&
  {Woitke}}{2006}]{Helling2006}
{Helling} C.,  {Woitke} P.,  2006, \mn@doi [\aap] {10.1051/0004-6361:20054598},
  \href {https://ui.adsabs.harvard.edu/abs/2006A&A...455..325H} {455, 325}

\bibitem[\protect\citeauthoryear{{Helling}, {Kawashima}, {Graham}, {Samra},
  {Chubb}, {Min}, {Waters}  \& {Parmentier}}{{Helling}
  et~al.}{2020}]{Helling2020}
{Helling} C.,  {Kawashima} Y.,  {Graham} V.,  {Samra} D.,  {Chubb} K.~L.,
  {Min} M.,  {Waters} L.~B.~F.~M.,   {Parmentier} V.,  2020, \mn@doi [\aap]
  {10.1051/0004-6361/202037633}, \href
  {https://ui.adsabs.harvard.edu/abs/2020A&A...641A.178H} {641, A178}

\bibitem[\protect\citeauthoryear{{Hohm}}{{Hohm}}{1994}]{Hohm1994}
{Hohm} U.,  1994, \mn@doi [Chemical Physics] {10.1016/0301-0104(94)87028-4},
  \href {https://ui.adsabs.harvard.edu/abs/1994CP....179..533H} {179, 533}

\bibitem[\protect\citeauthoryear{Hunter}{Hunter}{2007}]{Hunter:2007}
Hunter J.~D.,  2007, \mn@doi [Computing in Science \& Engineering]
  {10.1109/MCSE.2007.55}, 9, 90

\bibitem[\protect\citeauthoryear{{Jord{\'a}n} et~al.,}{{Jord{\'a}n}
  et~al.}{2013}]{Jordan2013}
{Jord{\'a}n} A.,  et~al., 2013, \mn@doi [\apj] {10.1088/0004-637X/778/2/184},
  \href {https://ui.adsabs.harvard.edu/abs/2013ApJ...778..184J} {778, 184}

\bibitem[\protect\citeauthoryear{{Karman} et~al.,}{{Karman}
  et~al.}{2019}]{Karman2019}
{Karman} T.,  et~al., 2019, \mn@doi [\icarus] {10.1016/j.icarus.2019.02.034},
  \href {https://ui.adsabs.harvard.edu/abs/2019Icar..328..160K} {328, 160}

\bibitem[\protect\citeauthoryear{{Kawashima} \& {Ikoma}}{{Kawashima} \&
  {Ikoma}}{2019}]{Kawashima2019}
{Kawashima} Y.,  {Ikoma} M.,  2019, \mn@doi [\apj] {10.3847/1538-4357/ab1b1d},
  \href {https://ui.adsabs.harvard.edu/abs/2019ApJ...877..109K} {877, 109}

\bibitem[\protect\citeauthoryear{{Kilpatrick} et~al.,}{{Kilpatrick}
  et~al.}{2018}]{KilpatrickEtal2018apjWASP63bWFC3}
{Kilpatrick} B.~M.,  et~al., 2018, \mn@doi [\aj] {10.3847/1538-3881/aacea7},
  \href {https://ui.adsabs.harvard.edu/abs/2018AJ....156..103K} {156, 103}

\bibitem[\protect\citeauthoryear{Kramida, {Yu.~Ralchenko}, Reader  \& {and NIST
  ASD Team}}{Kramida et~al.}{2018}]{Kramida2018}
Kramida A.,  {Yu.~Ralchenko} Reader J.,   {and NIST ASD Team} 2018, {NIST
  Atomic Spectra Database (ver. 5.6.1), [Online]. Available:
  {\tt{https://physics.nist.gov/asd}} [2019, February 6]. National Institute of
  Standards and Technology, Gaithersburg, MD.}

\bibitem[\protect\citeauthoryear{{Kreidberg} et~al.,}{{Kreidberg}
  et~al.}{2014}]{Kreidberg2014}
{Kreidberg} L.,  et~al., 2014, \mn@doi [\apjl]
  {10.1088/2041-8205/793/2/L2710.48550/arXiv.1410.2255}, \href
  {https://ui.adsabs.harvard.edu/abs/2014ApJ...793L..27K} {793, L27}

\bibitem[\protect\citeauthoryear{Lacis \& Oinas}{Lacis \&
  Oinas}{1991}]{Lacis1991}
Lacis A.~A.,  Oinas V.,  1991, Journal of Geophysical Research: Atmospheres,
  96, 9027

\bibitem[\protect\citeauthoryear{{Lavvas} \& {Koskinen}}{{Lavvas} \&
  {Koskinen}}{2017}]{Lavvas2017}
{Lavvas} P.,  {Koskinen} T.,  2017, \mn@doi [\apj] {10.3847/1538-4357/aa88ce},
  \href {https://ui.adsabs.harvard.edu/abs/2017ApJ...847...32L} {847, 32}

\bibitem[\protect\citeauthoryear{{Lecavelier Des Etangs}, {Pont},
  {Vidal-Madjar}  \& {Sing}}{{Lecavelier Des Etangs}
  et~al.}{2008}]{Lecavelier2008}
{Lecavelier Des Etangs} A.,  {Pont} F.,  {Vidal-Madjar} A.,   {Sing} D.,  2008,
  \mn@doi [\aap] {10.1051/0004-6361:200809388}, \href
  {https://ui.adsabs.harvard.edu/abs/2008A&A...481L..83L} {481, L83}

\bibitem[\protect\citeauthoryear{{Li}, {Gordon}, {Rothman}, {Tan}, {Hu},
  {Kassi}, {Campargue}  \& {Medvedev}}{{Li} et~al.}{2015}]{Li2015}
{Li} G.,  {Gordon} I.~E.,  {Rothman} L.~S.,  {Tan} Y.,  {Hu} S.-M.,  {Kassi}
  S.,  {Campargue} A.,   {Medvedev} E.~S.,  2015, \mn@doi [\apjs]
  {10.1088/0067-0049/216/1/15}, \href
  {https://ui.adsabs.harvard.edu/abs/2015ApJS..216...15L} {216, 15}

\bibitem[\protect\citeauthoryear{{Liddle}}{{Liddle}}{2007}]{Liddle2007}
{Liddle} A.~R.,  2007, \mn@doi [\mnras] {10.1111/j.1745-3933.2007.00306.x},
  \href {https://ui.adsabs.harvard.edu/abs/2007MNRAS.377L..74L} {377, L74}

\bibitem[\protect\citeauthoryear{{Line} \& {Parmentier}}{{Line} \&
  {Parmentier}}{2016}]{Line2016}
{Line} M.~R.,  {Parmentier} V.,  2016, \mn@doi [\apj]
  {10.3847/0004-637X/820/1/78}, \href
  {https://ui.adsabs.harvard.edu/abs/2016ApJ...820...78L} {820, 78}

\bibitem[\protect\citeauthoryear{{Line} et~al.,}{{Line}
  et~al.}{2013}]{Line2013}
{Line} M.~R.,  et~al., 2013, \mn@doi [\apj] {10.1088/0004-637X/775/2/137},
  \href {https://ui.adsabs.harvard.edu/abs/2013ApJ...775..137L} {775, 137}

\bibitem[\protect\citeauthoryear{{Line} et~al.,}{{Line}
  et~al.}{2016}]{Lineb2016}
{Line} M.~R.,  et~al., 2016, \mn@doi [\aj]
  {10.3847/0004-6256/152/6/20310.48550/arXiv.1605.08810}, \href
  {https://ui.adsabs.harvard.edu/abs/2016AJ....152..203L} {152, 203}

\bibitem[\protect\citeauthoryear{{Lodders} \& {Fegley}}{{Lodders} \&
  {Fegley}}{2002}]{Lodders2002}
{Lodders} K.,  {Fegley} B.,  2002, \mn@doi [\icarus] {10.1006/icar.2001.6740},
  \href {https://ui.adsabs.harvard.edu/abs/2002Icar..155..393L} {155, 393}

\bibitem[\protect\citeauthoryear{{MacDonald}}{{MacDonald}}{2023}]{MacDonald2023}
{MacDonald} R.~J.,  2023, \mn@doi [Journal of Open Source Software]
  {10.21105/joss.04873}, 8, 4873

\bibitem[\protect\citeauthoryear{{MacDonald} \& {Lewis}}{{MacDonald} \&
  {Lewis}}{2022}]{MacDonald2022}
{MacDonald} R.~J.,  {Lewis} N.~K.,  2022, \mn@doi [\apj]
  {10.3847/1538-4357/ac47fe}, \href
  {https://ui.adsabs.harvard.edu/abs/2022ApJ...929...20M} {929, 20}

\bibitem[\protect\citeauthoryear{{MacDonald} \& {Madhusudhan}}{{MacDonald} \&
  {Madhusudhan}}{2017}]{MacDonald2017}
{MacDonald} R.~J.,  {Madhusudhan} N.,  2017, \mn@doi [\mnras]
  {10.1093/mnras/stx804}, \href
  {https://ui.adsabs.harvard.edu/abs/2017MNRAS.469.1979M} {469, 1979}

\bibitem[\protect\citeauthoryear{{Madhusudhan}}{{Madhusudhan}}{2019}]{Madhusudhan2019}
{Madhusudhan} N.,  2019, \mn@doi [\araa] {10.1146/annurev-astro-081817-051846},
  \href {https://ui.adsabs.harvard.edu/abs/2019ARA&A..57..617M} {57, 617}

\bibitem[\protect\citeauthoryear{{Madhusudhan} \& {Seager}}{{Madhusudhan} \&
  {Seager}}{2009}]{Madhusudhan2009}
{Madhusudhan} N.,  {Seager} S.,  2009, \mn@doi [\apj]
  {10.1088/0004-637X/707/1/24}, \href
  {https://ui.adsabs.harvard.edu/abs/2009ApJ...707...24M} {707, 24}

\bibitem[\protect\citeauthoryear{{Malik} et~al.,}{{Malik}
  et~al.}{2017}]{MalikEtal2017ajHELIOS}
{Malik} M.,  et~al., 2017, \mn@doi [\aj] {10.3847/1538-3881/153/2/56}, \href
  {https://ui.adsabs.harvard.edu/abs/2017AJ....153...56M} {153, 56}

\bibitem[\protect\citeauthoryear{{McGruder} et~al.,}{{McGruder}
  et~al.}{2022}]{McGruder2022}
{McGruder} C.~D.,  et~al., 2022, \mn@doi [\aj] {10.3847/1538-3881/ac7f2e},
  \href {https://ui.adsabs.harvard.edu/abs/2022AJ....164..134M} {164, 134}

\bibitem[\protect\citeauthoryear{{Morley}, {Fortney}, {Marley}, {Visscher},
  {Saumon}  \& {Leggett}}{{Morley} et~al.}{2012}]{Morley2012}
{Morley} C.~V.,  {Fortney} J.~J.,  {Marley} M.~S.,  {Visscher} C.,  {Saumon}
  D.,   {Leggett} S.~K.,  2012, \mn@doi [\apj] {10.1088/0004-637X/756/2/172},
  \href {https://ui.adsabs.harvard.edu/abs/2012ApJ...756..172M} {756, 172}

\bibitem[\protect\citeauthoryear{{Nikolov} et~al.,}{{Nikolov}
  et~al.}{2018}]{Nikolov2018}
{Nikolov} N.,  et~al., 2018, \mn@doi [\nat] {10.1038/s41586-018-0101-7}, \href
  {https://ui.adsabs.harvard.edu/abs/2018Natur.557..526N} {557, 526}

\bibitem[\protect\citeauthoryear{{Nikolov} et~al.,}{{Nikolov}
  et~al.}{2022}]{Nikolov2022}
{Nikolov} N.~K.,  et~al., 2022, \mn@doi [\mnras] {10.1093/mnras/stac1530},
  \href {https://ui.adsabs.harvard.edu/abs/2022MNRAS.515.3037N} {515, 3037}

\bibitem[\protect\citeauthoryear{{Parmentier}, {Fortney}, {Showman}, {Morley}
  \& {Marley}}{{Parmentier} et~al.}{2016}]{Parmentier2016}
{Parmentier} V.,  {Fortney} J.~J.,  {Showman} A.~P.,  {Morley} C.,   {Marley}
  M.~S.,  2016, \mn@doi [\apj]
  {10.3847/0004-637X/828/1/2210.48550/arXiv.1602.03088}, \href
  {https://ui.adsabs.harvard.edu/abs/2016ApJ...828...22P} {828, 22}

\bibitem[\protect\citeauthoryear{{Parmentier} et~al.,}{{Parmentier}
  et~al.}{2018}]{2018A&A...617A.110P}
{Parmentier} V.,  et~al., 2018, \mn@doi [\aap] {10.1051/0004-6361/201833059},
  \href {https://ui.adsabs.harvard.edu/abs/2018A&A...617A.110P} {617, A110}

\bibitem[\protect\citeauthoryear{{Parmentier}, {Showman}  \&
  {Fortney}}{{Parmentier} et~al.}{2021}]{2021MNRAS.501...78P}
{Parmentier} V.,  {Showman} A.~P.,   {Fortney} J.~J.,  2021, \mn@doi [\mnras]
  {10.1093/mnras/staa3418}, \href
  {https://ui.adsabs.harvard.edu/abs/2021MNRAS.501...78P} {501, 78}

\bibitem[\protect\citeauthoryear{{Pinhas}, {Rackham}, {Madhusudhan}  \&
  {Apai}}{{Pinhas} et~al.}{2018}]{Pinhas2018}
{Pinhas} A.,  {Rackham} B.~V.,  {Madhusudhan} N.,   {Apai} D.,  2018, \mn@doi
  [\mnras] {10.1093/mnras/sty2209}, \href
  {https://ui.adsabs.harvard.edu/abs/2018MNRAS.480.5314P} {480, 5314}

\bibitem[\protect\citeauthoryear{{Polyansky}, {Kyuberis}, {Zobov}, {Tennyson},
  {Yurchenko}  \& {Lodi}}{{Polyansky} et~al.}{2018}]{Polyansky2018}
{Polyansky} O.~L.,  {Kyuberis} A.~A.,  {Zobov} N.~F.,  {Tennyson} J.,
  {Yurchenko} S.~N.,   {Lodi} L.,  2018, \mn@doi [\mnras]
  {10.1093/mnras/sty1877}, \href
  {https://ui.adsabs.harvard.edu/abs/2018MNRAS.480.2597P} {480, 2597}

\bibitem[\protect\citeauthoryear{{Pontoppidan} et~al.,}{{Pontoppidan}
  et~al.}{2022}]{Pontoppidan2022}
{Pontoppidan} K.~M.,  et~al., 2022, \mn@doi [\apjl] {10.3847/2041-8213/ac8a4e},
  \href {https://ui.adsabs.harvard.edu/abs/2022ApJ...936L..14P} {936, L14}

\bibitem[\protect\citeauthoryear{{Powell}, {Zhang}, {Gao}  \&
  {Parmentier}}{{Powell} et~al.}{2018}]{Powell2018}
{Powell} D.,  {Zhang} X.,  {Gao} P.,   {Parmentier} V.,  2018, \mn@doi [\apj]
  {10.3847/1538-4357/aac215}, \href
  {https://ui.adsabs.harvard.edu/abs/2018ApJ...860...18P} {860, 18}

\bibitem[\protect\citeauthoryear{{Rackham} et~al.,}{{Rackham}
  et~al.}{2017}]{Rackham2017}
{Rackham} B.,  et~al., 2017, \mn@doi [\apj] {10.3847/1538-4357/aa4f6c}, \href
  {https://ui.adsabs.harvard.edu/abs/2017ApJ...834..151R} {834, 151}

\bibitem[\protect\citeauthoryear{{Radica} et~al.,}{{Radica}
  et~al.}{2022}]{radica_applesoss}
{Radica} M.,  et~al., 2022, \mn@doi [\pasp] {10.1088/1538-3873/ac9430}, \href
  {https://ui.adsabs.harvard.edu/abs/2022PASP..134j4502R} {134, 104502}

\bibitem[\protect\citeauthoryear{Rajpurohit, Reyl{\'{e}}, Allard, Homeier,
  Schultheis, Bessell  \& Robin}{Rajpurohit et~al.}{2013}]{Rajpurohit2013}
Rajpurohit A.~S.,  Reyl{\'{e}} C.,  Allard F.,  Homeier D.,  Schultheis M.,
  Bessell M.~S.,   Robin A.~C.,  2013, \mn@doi [Astronomy and Astrophysics]
  {10.1051/0004-6361/201321346}, 556, 1

\bibitem[\protect\citeauthoryear{{Richard} et~al.,}{{Richard}
  et~al.}{2012}]{Richard2012}
{Richard} C.,  et~al., 2012, \mn@doi [\jqsrt] {10.1016/j.jqsrt.2011.11.004},
  \href {http://adsabs.harvard.edu/abs/2012JQSRT.113.1276R} {113, 1276}

\bibitem[\protect\citeauthoryear{{Rothman} et~al.,}{{Rothman}
  et~al.}{2010a}]{Rothman+10}
{Rothman} L.~S.,  et~al., 2010a, \mn@doi [\jqsrt]
  {10.1016/j.jqsrt.2010.05.001}, \href
  {https://ui.adsabs.harvard.edu/abs/2010JQSRT.111.2139R} {111, 2139}

\bibitem[\protect\citeauthoryear{{Rothman} et~al.,}{{Rothman}
  et~al.}{2010b}]{Rothman2010}
{Rothman} L.~S.,  et~al., 2010b, \mn@doi [\jqsrt]
  {10.1016/j.jqsrt.2010.05.001}, \href
  {https://ui.adsabs.harvard.edu/abs/2010JQSRT.111.2139R} {111, 2139}

\bibitem[\protect\citeauthoryear{{Rustamkulov} et~al.,}{{Rustamkulov}
  et~al.}{2022}]{Rustamkulov2022}
{Rustamkulov} Z.,  et~al., 2022, arXiv e-prints, \href
  {https://ui.adsabs.harvard.edu/abs/2022arXiv221110487R} {p. arXiv:2211.10487}

\bibitem[\protect\citeauthoryear{{Ryabchikova}, {Piskunov}, {Kurucz},
  {Stempels}, {Heiter}, {Pakhomov}  \& {Barklem}}{{Ryabchikova}
  et~al.}{2015}]{Ryabchikova2015}
{Ryabchikova} T.,  {Piskunov} N.,  {Kurucz} R.~L.,  {Stempels} H.~C.,  {Heiter}
  U.,  {Pakhomov} Y.,   {Barklem} P.~S.,  2015, \mn@doi [Physica Scripta]
  {10.1088/0031-8949/90/5/054005}, \href
  {https://ui.adsabs.harvard.edu/abs/2015PhyS...90e4005R} {90, 054005}

\bibitem[\protect\citeauthoryear{{Samra}, {Helling}, {Chubb}, {Min}, {Carone}
  \& {Schneider}}{{Samra} et~al.}{2023}]{Samra2022}
{Samra} D.,  {Helling} C.,  {Chubb} K.~L.,  {Min} M.,  {Carone} L.,
  {Schneider} A.~D.,  2023, \mn@doi [\aap] {10.1051/0004-6361/202244939}, \href
  {https://ui.adsabs.harvard.edu/abs/2023A&A...669A.142S} {669, A142}

\bibitem[\protect\citeauthoryear{{Showman}, {Fortney}, {Lian}, {Marley},
  {Freedman}, {Knutson}  \& {Charbonneau}}{{Showman}
  et~al.}{2009}]{2009ApJ...699..564S}
{Showman} A.~P.,  {Fortney} J.~J.,  {Lian} Y.,  {Marley} M.~S.,  {Freedman}
  R.~S.,  {Knutson} H.~A.,   {Charbonneau} D.,  2009, \mn@doi [\apj]
  {10.1088/0004-637X/699/1/564}, \href
  {https://ui.adsabs.harvard.edu/abs/2009ApJ...699..564S} {699, 564}

\bibitem[\protect\citeauthoryear{{Showman}, {Tan}  \& {Parmentier}}{{Showman}
  et~al.}{2020}]{Showman2020}
{Showman} A.~P.,  {Tan} X.,   {Parmentier} V.,  2020, \mn@doi [\ssr]
  {10.1007/s11214-020-00758-8}, \href
  {https://ui.adsabs.harvard.edu/abs/2020SSRv..216..139S} {216, 139}

\bibitem[\protect\citeauthoryear{{Steinrueck}, {Showman}, {Lavvas}, {Koskinen},
  {Tan}  \& {Zhang}}{{Steinrueck} et~al.}{2021}]{Steinrueck2021}
{Steinrueck} M.~E.,  {Showman} A.~P.,  {Lavvas} P.,  {Koskinen} T.,  {Tan} X.,
   {Zhang} X.,  2021, \mn@doi [\mnras] {10.1093/mnras/stab1053}, \href
  {https://ui.adsabs.harvard.edu/abs/2021MNRAS.504.2783S} {504, 2783}

\bibitem[\protect\citeauthoryear{{Tashkun} \& {Perevalov}}{{Tashkun} \&
  {Perevalov}}{2011}]{Tashkun2011}
{Tashkun} S.~A.,  {Perevalov} V.~I.,  2011, \mn@doi [Journal of Quantitative
  Spectroscopy and Radiative Transfer] {10.1016/j.jqsrt.2011.03.005}, \href
  {https://ui.adsabs.harvard.edu/\#abs/2011JQSRT.112.1403T} {112, 1403}

\bibitem[\protect\citeauthoryear{{Tennyson} et~al.,}{{Tennyson}
  et~al.}{2016}]{TennysonEtal2016jmsExomol}
{Tennyson} J.,  et~al., 2016, \mn@doi [Journal of Molecular Spectroscopy]
  {10.1016/j.jms.2016.05.002}, \href
  {https://ui.adsabs.harvard.edu/abs/2016JMoSp.327...73T} {327, 73}

\bibitem[\protect\citeauthoryear{{The JWST Transiting Exoplanet Community Early
  Release Science Team} et~al.,}{{The JWST Transiting Exoplanet Community Early
  Release Science Team} et~al.}{2022}]{ERS2022}
{The JWST Transiting Exoplanet Community Early Release Science Team} et~al.,
  2022, arXiv e-prints, \href
  {https://ui.adsabs.harvard.edu/abs/2022arXiv220811692T} {p. arXiv:2208.11692}

\bibitem[\protect\citeauthoryear{{Tsai} et~al.,}{{Tsai}
  et~al.}{2022}]{Tsai2022}
{Tsai} S.-M.,  et~al., 2022, arXiv e-prints, \href
  {https://ui.adsabs.harvard.edu/abs/2022arXiv221110490T} {p. arXiv:2211.10490}

\bibitem[\protect\citeauthoryear{Venot, H{\'{e}}brard, Ag{\'{u}}ndez,
  Dobrijevic, Selsis, Hersant, Iro  \& Bounaceur}{Venot
  et~al.}{2012}]{Venot2012}
Venot O.,  H{\'{e}}brard E.,  Ag{\'{u}}ndez M.,  Dobrijevic M.,  Selsis F.,
  Hersant F.,  Iro N.,   Bounaceur R.,  2012, \mn@doi [Astronomy and
  Astrophysics] {10.1051/0004-6361/201219310}, 546

\bibitem[\protect\citeauthoryear{Venot, Bounaceur, Dobrijevic, H{\'{e}}brard,
  Cavali{\'{e}}, Tremblin, Drummond  \& Charnay}{Venot
  et~al.}{2019}]{Venot2019}
Venot O.,  Bounaceur R.,  Dobrijevic M.,  H{\'{e}}brard E.,  Cavali{\'{e}} T.,
  Tremblin P.,  Drummond B.,   Charnay B.,  2019, \mn@doi [Astronomy and
  Astrophysics] {10.1051/0004-6361/201834861}, 624, 1

\bibitem[\protect\citeauthoryear{{Venot} et~al.,}{{Venot}
  et~al.}{2020}]{VenotEtal2020-JWST-WASP-43b}
{Venot} O.,  et~al., 2020, \mn@doi [\apj] {10.3847/1538-4357/ab6a94}, \href
  {https://ui.adsabs.harvard.edu/abs/2020ApJ...890..176V} {890, 176}

\bibitem[\protect\citeauthoryear{Virtanen et~al.,}{Virtanen
  et~al.}{2020}]{2020SciPy-NMeth}
Virtanen P.,  et~al., 2020, \mn@doi [Nature Methods]
  {10.1038/s41592-019-0686-2}, \href {https://rdcu.be/b08Wh} {17, 261}

\bibitem[\protect\citeauthoryear{{Welbanks} \& {Madhusudhan}}{{Welbanks} \&
  {Madhusudhan}}{2019}]{Welbanks2019a}
{Welbanks} L.,  {Madhusudhan} N.,  2019, \mn@doi [\aj]
  {10.3847/1538-3881/ab14de}, \href
  {https://ui.adsabs.harvard.edu/abs/2019AJ....157..206W} {157, 206}

\bibitem[\protect\citeauthoryear{{Welbanks} \& {Madhusudhan}}{{Welbanks} \&
  {Madhusudhan}}{2021}]{Welbanks2021}
{Welbanks} L.,  {Madhusudhan} N.,  2021, \mn@doi [\apj]
  {10.3847/1538-4357/abee94}, \href
  {https://ui.adsabs.harvard.edu/abs/2021ApJ...913..114W} {913, 114}

\bibitem[\protect\citeauthoryear{{Welbanks} \& {Madhusudhan}}{{Welbanks} \&
  {Madhusudhan}}{2022}]{Welbanks2022}
{Welbanks} L.,  {Madhusudhan} N.,  2022, \mn@doi [\apj]
  {10.3847/1538-4357/ac6df1}, \href
  {https://ui.adsabs.harvard.edu/abs/2022ApJ...933...79W} {933, 79}

\bibitem[\protect\citeauthoryear{{Welbanks}, {Madhusudhan}, {Allard}, {Hubeny},
  {Spiegelman}  \& {Leininger}}{{Welbanks} et~al.}{2019}]{Welbanks2019}
{Welbanks} L.,  {Madhusudhan} N.,  {Allard} N.~F.,  {Hubeny} I.,  {Spiegelman}
  F.,   {Leininger} T.,  2019, \mn@doi [\apjl] {10.3847/2041-8213/ab5a89},
  \href {https://ui.adsabs.harvard.edu/abs/2019ApJ...887L..20W} {887, L20}

\bibitem[\protect\citeauthoryear{{Yip}, {Changeat}, {Edwards}, {Morvan},
  {Chubb}, {Tsiaras}, {Waldmann}  \& {Tinetti}}{{Yip} et~al.}{2021}]{Yip2021}
{Yip} K.~H.,  {Changeat} Q.,  {Edwards} B.,  {Morvan} M.,  {Chubb} K.~L.,
  {Tsiaras} A.,  {Waldmann} I.~P.,   {Tinetti} G.,  2021, \mn@doi [\aj]
  {10.3847/1538-3881/abc179}, \href
  {https://ui.adsabs.harvard.edu/abs/2021AJ....161....4Y} {161, 4}

\bibitem[\protect\citeauthoryear{{Yurchenko}}{{Yurchenko}}{2015}]{YurchenkoEtal2015}
{Yurchenko} S.~N.,  2015, \mn@doi [\jqsrt] {10.1016/j.jqsrt.2014.10.023}, \href
  {https://ui.adsabs.harvard.edu/abs/2015JQSRT.152...28Y} {152, 28}

\bibitem[\protect\citeauthoryear{{Yurchenko} \& {Tennyson}}{{Yurchenko} \&
  {Tennyson}}{2014}]{Yurchenko2014}
{Yurchenko} S.~N.,  {Tennyson} J.,  2014, \mn@doi [\mnras]
  {10.1093/mnras/stu326}, \href
  {https://ui.adsabs.harvard.edu/abs/2014MNRAS.440.1649Y} {440, 1649}

\bibitem[\protect\citeauthoryear{{Yurchenko}, {Barber}  \&
  {Tennyson}}{{Yurchenko} et~al.}{2011}]{YurchenkoEtal2011}
{Yurchenko} S.~N.,  {Barber} R.~J.,   {Tennyson} J.,  2011, \mn@doi [\mnras]
  {10.1111/j.1365-2966.2011.18261.x}, \href
  {https://ui.adsabs.harvard.edu/abs/2011MNRAS.413.1828Y} {413, 1828}

\bibitem[\protect\citeauthoryear{{Yurchenko}, {Amundsen}, {Tennyson}  \&
  {Waldmann}}{{Yurchenko} et~al.}{2017}]{Yurchenko2017}
{Yurchenko} S.~N.,  {Amundsen} D.~S.,  {Tennyson} J.,   {Waldmann} I.~P.,
  2017, \mn@doi [Astronomy and Astrophysics] {10.1051/0004-6361/201731026},
  \href {https://ui.adsabs.harvard.edu/\#abs/2017Astronomy and
  Astrophysics...605A..95Y} {605, A95}

\bibitem[\protect\citeauthoryear{Zamyatina et~al.,}{Zamyatina
  et~al.}{2023}]{Zamyatina2023}
Zamyatina M.,  et~al., 2023, \mn@doi [Monthly Notices of the Royal Astronomical
  Society] {10.1093/mnras/stac3432}, 519, 3129

\makeatother
\end{thebibliography}



\appendix

\section{Posterior distribution for final model}
\label{sec: posteriors}

\begin{figure*}
    \centering
    \includegraphics[width=\textwidth]{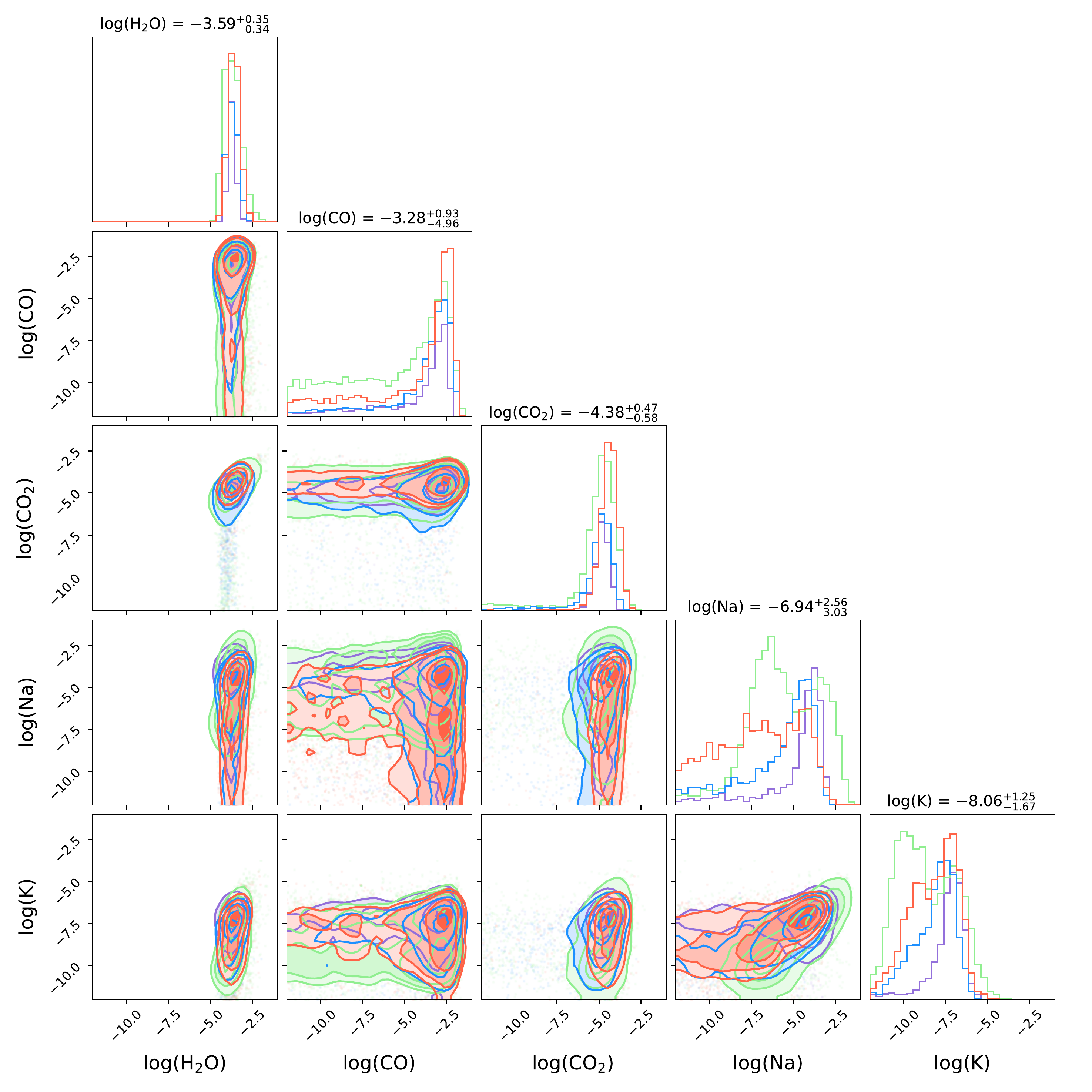}
    \caption{The posterior distribution of the key atmospheric consitutiants. The colour scheme is the same as Figure \ref{fig:WASP-96b_Results} (CHIMERA = purple, Pyratbay = green, POSEIDON = blue, and Aurora = red), with the reference quoted abundance constraints above each histogram being from the Aurora framework.}
    \label{fig:cornerplot}
\end{figure*}

\begin{figure*}
    \centering
    \includegraphics[width=\textwidth]{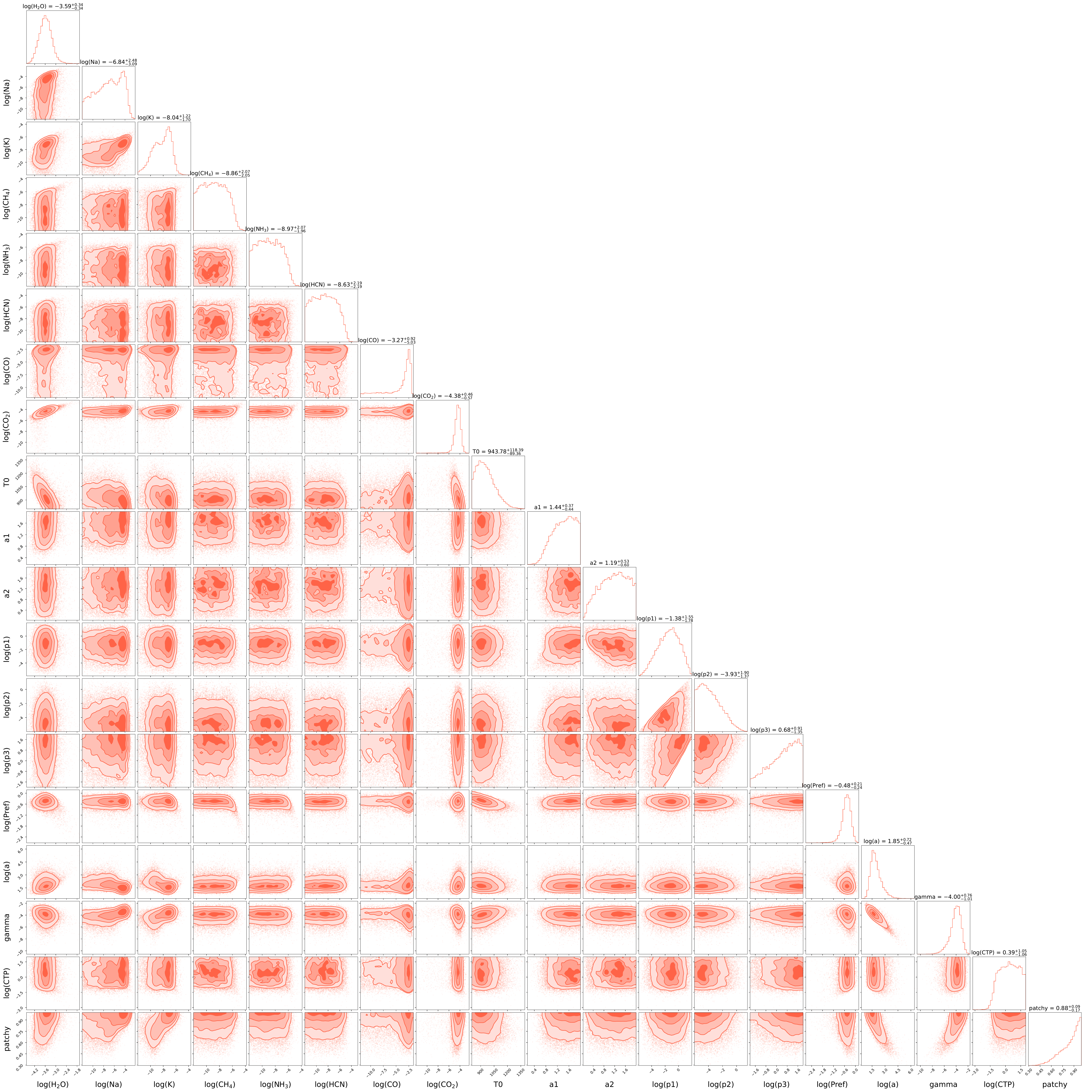}
    \caption{The posterior distribution from the Aurora framework.}
    \label{fig:aurora_corner}
\end{figure*}

\begin{figure*}
    \centering
    \includegraphics[width=\textwidth]{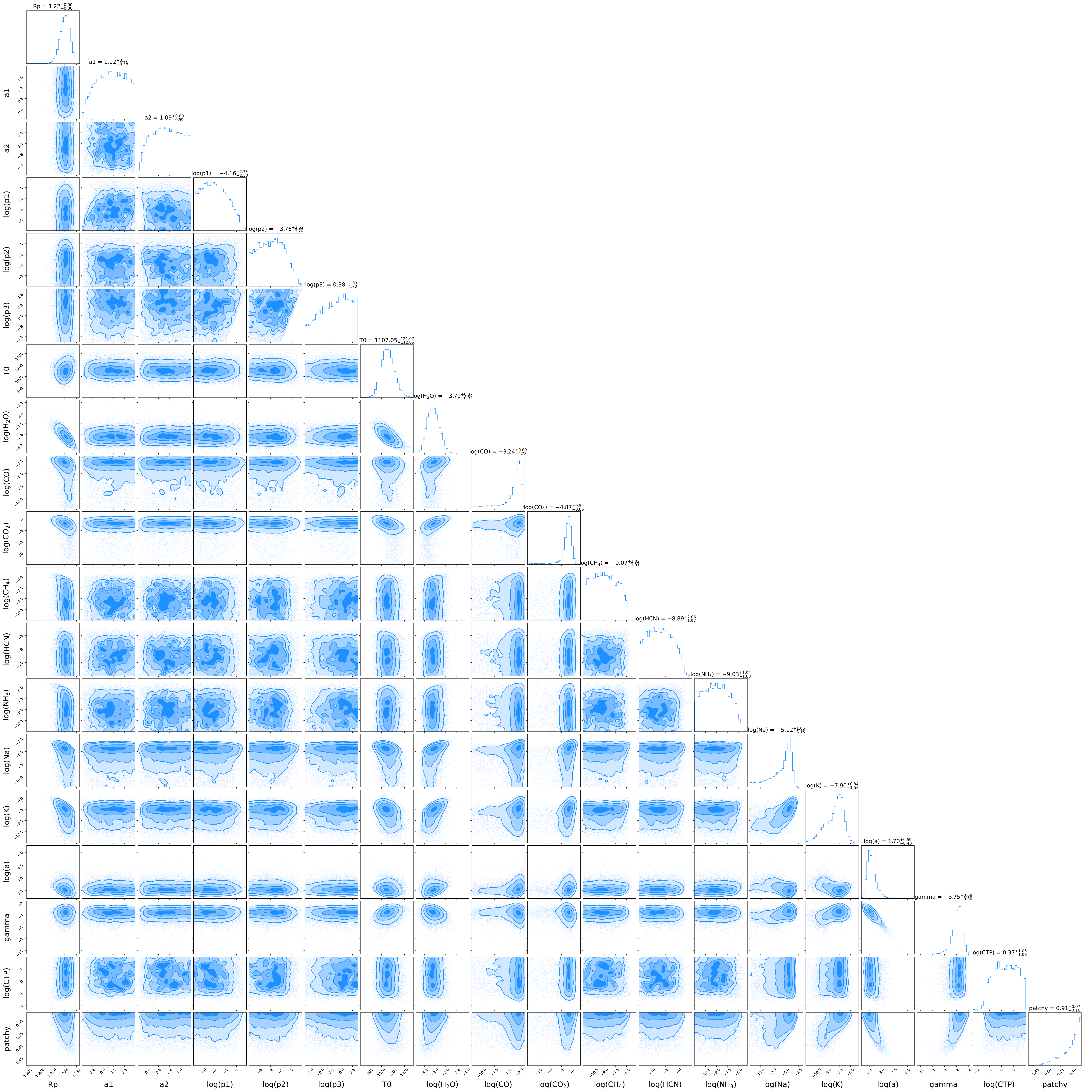}
    \caption{The posterior distribution from the POSEIDON framework.}
    \label{fig:poseidon_corner}
\end{figure*}

\begin{figure*}
    \centering
    \includegraphics[width=\textwidth]{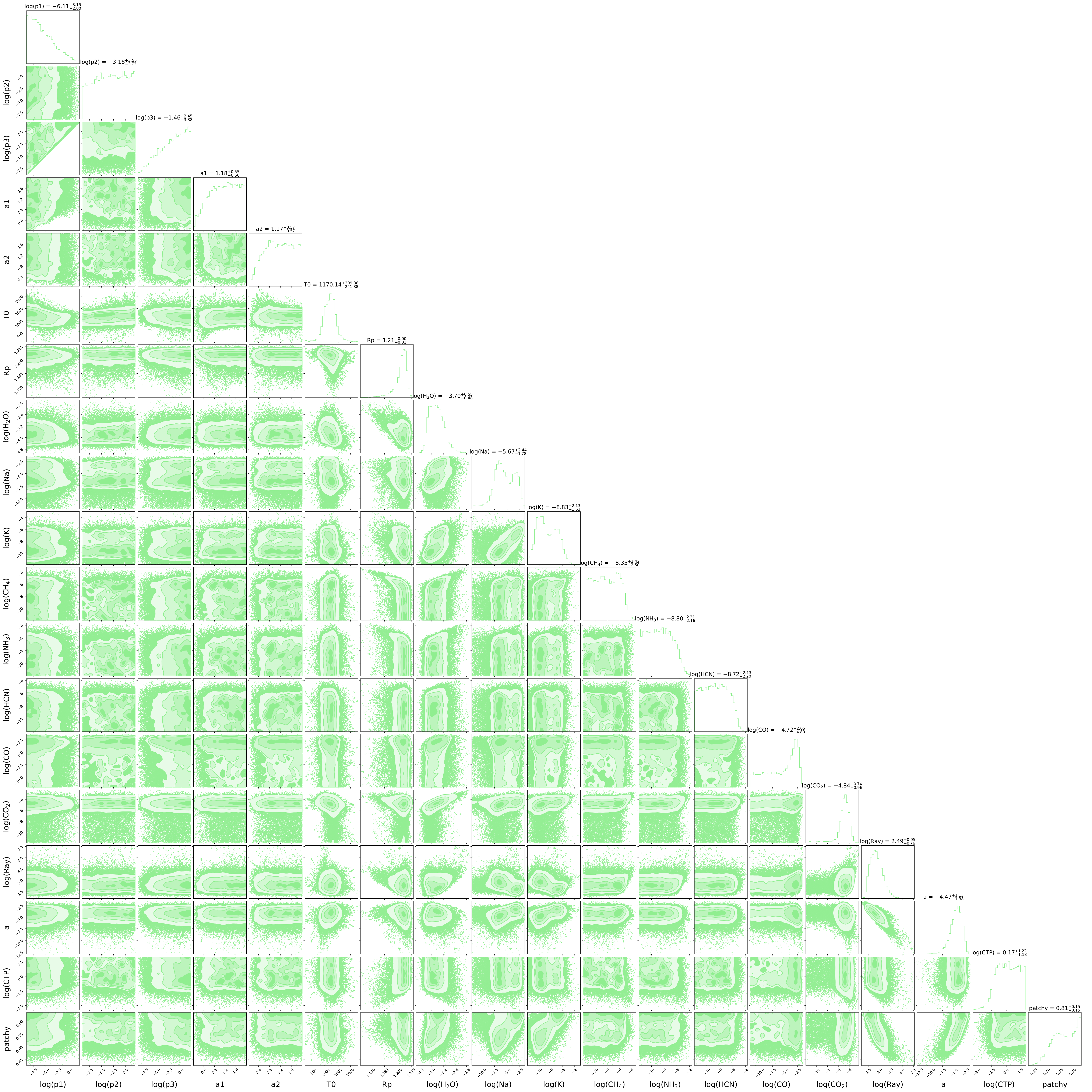}
    \caption{The posterior distribution from the PyratBay framework.}
    \label{fig:pyratbay_corner}
\end{figure*}

\begin{figure*}
    \centering
    \includegraphics[width=\textwidth]{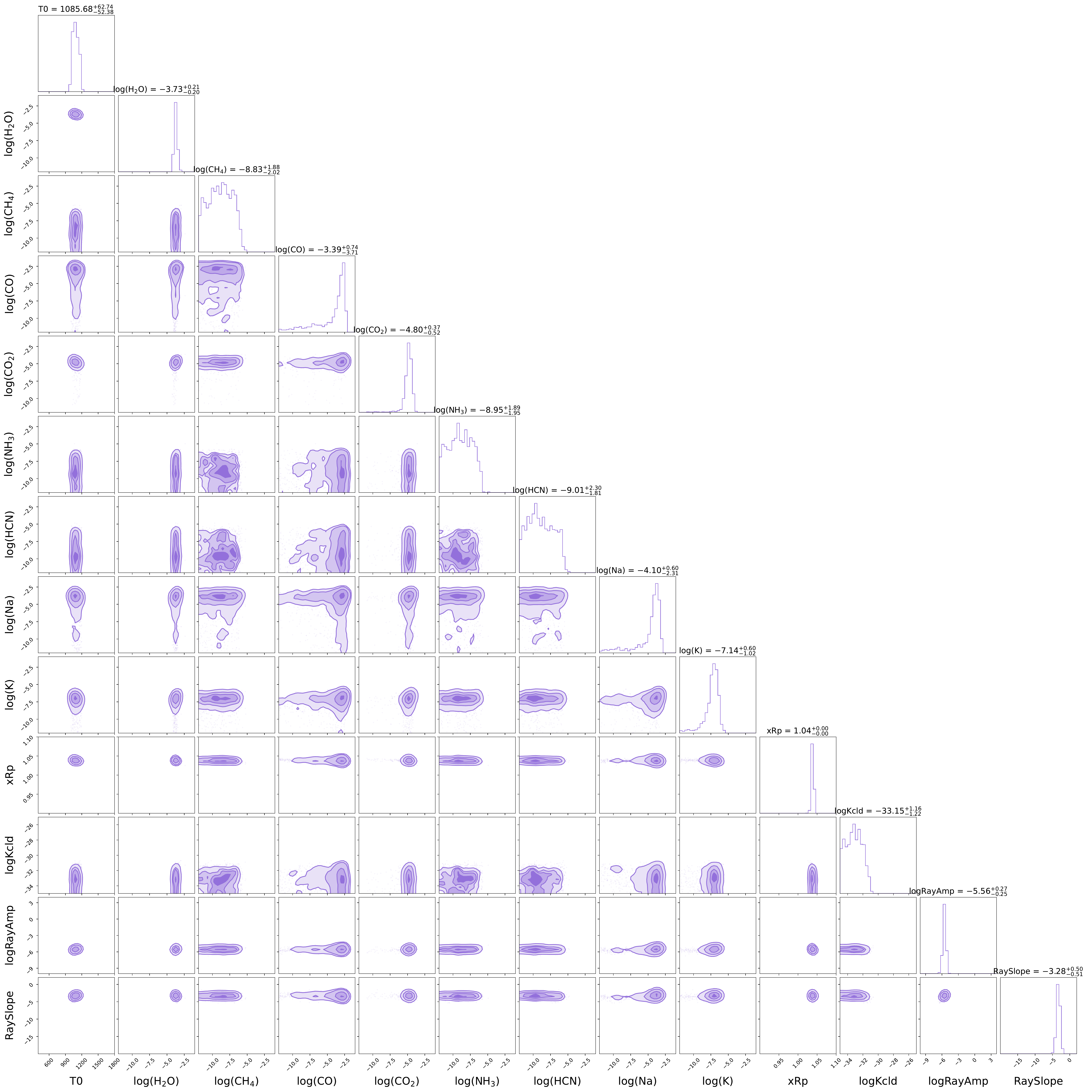}
    \caption{The posterior distribution from the CHIMERA framework.}
    \label{fig:chimera_corner}
\end{figure*}

\begin{figure*}
    \centering
    \includegraphics[width=\textwidth]{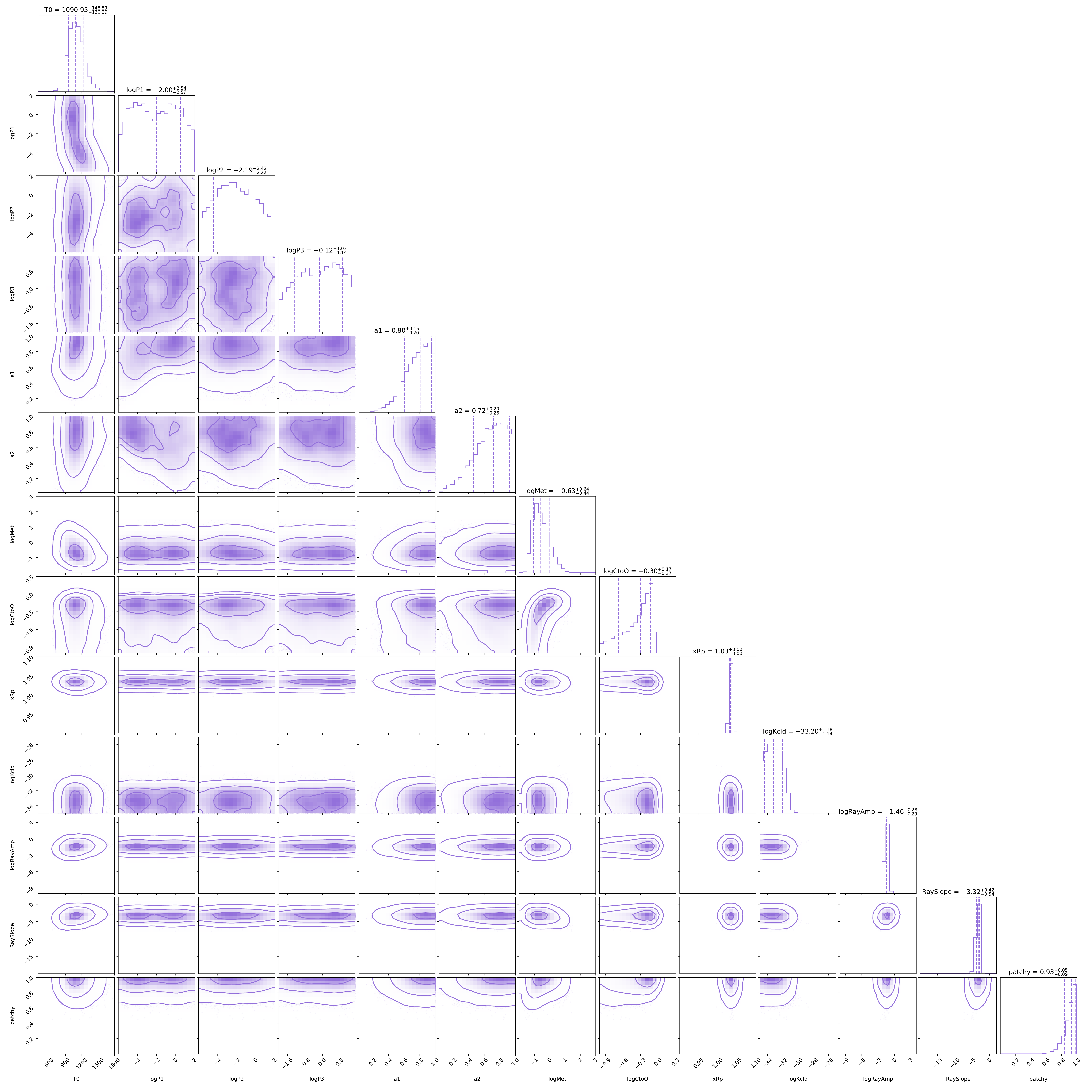}
    \caption{The chemical equilibrium posterior distribution from the CHIMERA framework.}
    \label{fig:chimera_corner_CE}
\end{figure*}



\bsp	
\label{lastpage}
\end{document}